\documentclass[12pt]{article} %\input epsf.tex

\usepackage{cite}
\usepackage{graphicx,subfigure,latexsym,amssymb}
\usepackage{bm}
\usepackage{amsmath}
\usepackage{color}
\usepackage{hyperref}

\usepackage{tikz}
\usepackage{tikz-feynhand}
\usepackage[normalem]{ulem}  % \sout{old text} for strikeout

\hypersetup{% hyperref option list
setpagesize=false,
 bookmarksnumbered=true,%
 bookmarksopen=true,%
}

\usepackage{chngcntr}
\counterwithin{equation}{section}

\newcommand{\be}{\begin{equation}}
\newcommand{\ee}{\end{equation}}
\newcommand{\bear}{\begin{eqnarray}}
\newcommand{\eear}{\end{eqnarray}}
\newcommand{\ba}{\begin{array}}
\newcommand{\ea}{\end{array}}

\newcommand{\cout}[1]{ \if 0 {#1} \fi }

\renewcommand{\thefootnote}{\fnsymbol{footnote}}
\begin{document}

\begin{titlepage}
%\vfill
%\begin{flushright}
%{\normalsize }\\
%\end{flushright}

\vfill
\begin{center}
{\Large\bf  Schwinger-Keldysh effective action for hydrodynamics with approximate symmetries}

\vskip 0.3in

Masaru Hongo$^{1,2}$\footnote{e-mail:
{\tt hongo@phys.sc.niigata-u.ac.jp}}, 
Noriyuki Sogabe$^{3}$\footnote{e-mail:
{\tt sogabe@uic.edu}}, 
Mikhail A. Stephanov$^{3}$\footnote{e-mail:
{\tt misha@uic.edu}}, 
Ho-Ung Yee$^{3}$\footnote{e-mail:
{\tt hyee@uic.edu}}
\vskip 0.3in

{\it $^{1}$ Department of Physics, Niigata University, Niigata 950-2181, Japan}
\\[0.15in]
{\it $^{2}$ RIKEN iTHEMS, RIKEN, Wako 351-0198, Japan}\\[0.15in]
{\it $^{3}$ Department of Physics, University of Illinois, Chicago, Illinois 60607}
\\[0.3in]

\end{center}

\begin{abstract}

We study the hydrodynamic theories with approximate symmetries in the recently developed effective action approach on the Schwinger-Keldysh (SK) contour. 
We employ the method of spurious symmetry transformation for small explicit symmetry-breaking parameters to systematically constrain symmetry-breaking effects in the non-equilibrium effective action for hydrodynamics. We apply our method to the hydrodynamic theory of chiral symmetry in Quantum Chromodynamics (QCD) at finite temperature and density and its explicit breaking by quark masses. We show that the spurious symmetry and the Kubo-Martin-Schwinger (KMS) relation dictate that the Ward-Takahashi identity for the axial symmetry, i.e., the partial conservation of axial vector current (PCAC) relation, contains a relaxational term proportional to the axial chemical potential, whose kinetic coefficient is at least of the second order in the quark mass. 
In the phase where the chiral symmetry is spontaneously broken, and the pseudo-Nambu-Goldstone pions appear as hydrodynamic variables, this relaxation effect is subleading compared to the conventional pion mass term in the PCAC relation, which is of the first order in the quark mass. 
On the other hand, in the chiral symmetry-restored phase, we show that our relaxation term, which is of the second order in the quark mass, becomes the leading contribution to the axial charge relaxation. 
Therefore, the leading axial charge relaxation mechanism is parametrically different in the quark mass across a chiral phase transition.

\end{abstract}

\vfill

\end{titlepage}
\renewcommand{\thefootnote}{\arabic{footnote}}  % 数字に戻す
\setcounter{footnote}{0} 

\section{Introduction \label{sec1}}

Symmetries, even if approximate, provide powerful constraints on both static and dynamic properties of a system. 
One of the most well-known examples in high-energy physics is the chiral symmetry of Quantum Chromodynamics (QCD), ${\rm SU}(2)_{\rm L}\times {\rm SU}(2)_{\rm R}$, which is explicitly broken by small but nonvanishing quark masses $m_{\rm q}$, more precisely a quark mass matrix $M={\rm diag}(m_{\rm u}, m_{\rm d})$.
In the QCD vacuum, the chiral symmetry is further spontaneously broken by the quark--anti-quark condensate $\left< \bar q q\right>\neq 0 $ even if there is no explicit breaking, $m_{\rm q}=0$.
As a consequence, the Nambu-Goldstone theorem~\cite{Nambu:1961tp,Goldstone:1961eq,Goldstone:1962es} dictates the appearence of pseudo-Nambu-Goldstone (NG) bosons, which are identified as pions.

The low-energy dynamics of pions is described by an effective field theory (EFT) constrained by the chiral symmetry and its explicit breaking by small quark masses; the chiral perturbation theory \cite{Weinberg:1978kz,Gasser:1983yg,Gasser:1984gg}. 
A convenient concept used in its construction is the spurious symmetry transformation acting on the quark mass matrix (see, e.g., Refs.~\cite{Georgi:1984zwz,Scherer-Schindler2011}), which, together with the original symmetry transformation on dynamical fields, defines the enlarged symmetry of the EFT. 
Combined with the appropriate power counting, requiring the EFT to be invariant under the enlarged symmetry transformation systematically incorporates the explicit symmetry-breaking effects as perturbations.

At a finite temperature $T$ that is low enough so that $\left< \bar q q\right> \neq 0$, the pions must be included as one of the hydrodynamic variables in the chiral limit $m_{\mathrm{q}} = 0$. 
In the presence of small quark masses, the chiral symmetry becomes approximate, and the pions have finite correlation time and length, which is of order $m_\pi^{-1}$. 
Thus, the validity of such hydrodynamics with pion degrees of freedom is restricted to a coressponding quasi-hydrodynamic or Hydro+ regime~\cite{Stephanov:2017ghc}, where time and length scales sufficiently shorter than $m_\pi^{-1}$. 
Beyond this scale, both the pions and the axial charges, which are inseparable from each other, become averaged out and disappear from the hydrodynamics description.

The hydrodynamics description of pions has been developed for over two decades, beginning with the pioneering work in the chiral and dissipation-less limits \cite{Son:1999pa}. 
Subsequently, it has been extended to include the dissipative and non-zero quark mass effects~\cite{Son:2002ci}. 
More recently, the second law of thermodynamics \cite{Grossi:2020ezz} was used to show a constraint between two kinetic coefficients introduced in Ref.~\cite{Son:2002ci}. 
These works assume the partial conservation of axial current (PCAC) relation~\cite{Nambu:1960xd,Gell-Mann:1960mvl} to the first order in quark mass,
\begin{align}
\label{eq:PCAC}
\partial_\mu j_{{\rm A}a}^{\mu} = \left< \bar q q\right> m_{\rm q} \pi^a \,, 
\end{align}
where $j_{{\rm A}a}^{\mu} \equiv \bar q \gamma^\mu \gamma_5 (\sigma^a/2) q$ is the axial isospin current with $\sigma^a$ being the Pauli matrices and $\pi^a$ denotes the pion field. However, the PCAC relation (\ref{eq:PCAC}) may, in general, be modified by dissipative effects, which have not been studied systematically. 
This is because
the source term [the right-hand side of (\ref{eq:PCAC})] can be seen as a 
constitutive relation from the viewpoint of hydrodynamics, which can generally be written in terms of hydrodynamic variables of the system, i.e., $j_{\rm Aa}^0$ and $\pi^a$.
It can, for example, contain a linear term proportional to the axial isospin charge density (or the corresponding chemical potential) causing the damping of the charge densities.

Recently, there have been theoretical developments to formulate dissipative hydrodynamics based on the EFT on the Schwinger-Keldysh (SK) contour \cite{Crossley:2015,Glorioso:2017fpd} (see also Refs.~\cite{Endlich:2012vt,Grozdanov:2013dba,Kovtun:2014hpa,Haehl:2015foa,Jensen:2017kzi,Glorioso:2018wxw,Chen-Lin:2018kfl,Landry:2019iel,Landry:2020ire,Baggioli:2023tlc,Akyuz:2023lsm,Huang:2024rml,Ota:2024yws} for complementary or further works and Refs.~\cite{Minami:2015uzo,Sieberer:2015svu,Hongo:2018ant,Hongo:2019qhi,Hidaka:2019irz,Li2024} for extentions to nonequilibrium open systems). 
These nonequilibrium EFTs are based on the action principle together with the symmetries of a microscopic theory on the SK contour, which provides an efficient framework for systematic construction and analysis of hydrodynamics relying on derivative expansion.
In particular, Ref.~\cite{Glorioso:2020loc} applied this framework to describe a lattice system with non-Abelian symmetries in the symmetry-restored phase. 
Moreover, Ref.~\cite{Delacretaz:2021qqu} explored relaxation dynamics of pseudo-NG modes in the symmetry-broken phase at the linearized level of the NG field.

In the nonequilibrium EFTs, the explicit symmetry-breaking terms are introduced by writing down all possible terms that break the symmetry at the lowest order of the dynamical fields (see e.g., Ref.~\cite{Delacretaz:2021qqu} and Refs.~\cite{Sogabe:2021wqk,Abbasi:2022aao}, for symmetry-broken and symmetry-restored phases, respectively). 
More specifically, this approach introduces a relaxation (or damping) term in the PCAC relation that is linear in the chemical potential $\mu_{{\rm A}a}$ corresponding to the axial charge $j^0_{\rm Aa}$,
\begin{align}
\label{eq:PCAC-naive}
\partial_\mu j_{{\rm A}a}^{\mu} = \left< \bar q q\right> m_{\rm q} \pi^a - \Gamma_{\rm A} \mu_{{\rm A}a}\,.
\end{align}
Here, $\Gamma_{\rm A}$ is a kinetic coefficient responsible for the axial charge damping. However, from the perspective of spurious symmetry transformation, we need a more systematic approach that ensures the invariance of the EFT action under the enlarged symmetry, where the explicit breaking parameter is elevated to a dynamical field that also transforms under the (spurious) symmetry group. 
Without such a systematic approach, we cannot determine the quark mass dependence of $\Gamma_{\rm A}=\Gamma_{\rm A}(m_{\rm q})$, other than the obvious expectation that $\Gamma_{\rm A}$ vanishes when $m_{\rm q}=0$.  
Thus, in order to formulate the EFT for hydrodynamics associated with the approximate chiral symmetry of QCD, we aim to construct the fully non-linear EFT action for a spontaneously broken non-Abelian symmetry while keeping the spurious symmetry.

The main objective of this work is to apply the idea of spurious symmetry to the construction of effective action for dissipative hydrodynamics with the approximate chiral symmetry of QCD at finite temperature and density. Based on the spurious chiral symmetry together with the so-called Kubo-Martin-Schwinger (KMS) symmetry that ensures thermal equilibrium satisfying the fluctuation-dissipation theorem, we show that the relaxation term highlighted above should be at least of second order in the quark mass, i.e., the coefficient $\Gamma_{\rm A}$ in Eq.~(\ref{eq:PCAC-naive}) must have the following dependence on the quark mass in lowest order perturbative expansion, 
\begin{align}
\label{eq:Gamma-mq}
\Gamma_{\rm A} = \gamma_{\rm A} m_{\rm q}^2\,.
\end{align}
We note that $\gamma_{\rm A}$ is nonvanishing in the chiral limit and
its dependence on other parameters, such as $T$, is determined by the details of the microscopic theory. 
This indicates that the PCAC relation is still a good approximation, even in a (quasi-)hydrodynamic regime, to the leading order of the quark mass, as
\begin{align}
\label{eq:PCAC-final}
\partial_\mu j_{{\rm A}a}^{\mu} = \left< \bar q q\right> m_{\rm q} \pi^a + \mathcal O(m_{\rm q}^2) \quad (\text{chiral\ SSB\ phase})\,.
\end{align}
where the relaxation term introduced in Eq.~(\ref{eq:PCAC-naive}) is systematically a higher-order term of the order $\mathcal O(m_{\rm q}^2)$.

On the other hand, our framework is equally applicable to the symmetry-restored phase, where the same parametric dependence on the quark mass (\ref{eq:Gamma-mq}) holds for $\Gamma_{\rm A}$. Furthermore, it becomes the leading contribution to the axial charge relaxation because the order parameter $\left< \bar q q\right>$ vanishes in this phase, 
\begin{align}
\label{eq:dj-restored}
\partial_\mu j_{{\rm A}a}^{\mu} = - \Gamma_{\rm A} \mu_{{\rm A}a}\quad (\text{chiral symmetry restored\  phase})\,.
\end{align}
In fact, it has been argued in the weakly coupled regime of high-temperature QCD that $\Gamma_{\rm A} \sim \alpha_s m_{\rm q}^2 T$ \cite{Hou:2017szz} with $\alpha_s=g^2/(4\pi)$ being the QCD coupling constant. Therefore, we conclude that the leading axial charge relaxation mechanism is parametrically different in the quark mass across a chiral transition between the symmetry-broken and the symmetry-restored phases.

Our method and the results are general enough to be applicable to any system with approximate internal symmetries, where spurious symmetry transformations can be introduced to the symmetry-breaking parameters. 
We will discuss the examples with ${\rm U}(1)$ and ${\rm SU}(N)$, and eventually approximate ${\rm SU}(N)_{\rm L}\times {\rm SU}(N)_{\rm R}$ chiral symmetry in QCD. 
In fact, the quadratic dependence of the charge damping rate on the explicit symmetry breaking parameter (\ref{eq:Gamma-mq}) has been widely observed in other fermion systems without SSB, such as supernovae and neutron star plasma \cite{PhysRevD.91.085035} and electron-positron plasma \cite{Boyarsky:2020ani}, etc., where the axial charge relaxation is caused by the chirality (or helicity) flipping prcesses.

The outline of the paper is as follows. In Sec.~\ref{sec:EFT-review}, we review the construction of the low-energy effective theory on the SK contour for non-Abelian symmetries. 
In Sec.~\ref{sec:U(1)-warm-up}, we start with an approximate ${\rm U}(1)$ symmetry as a simple warm-up and introduce the spurious symmetry transformation for the explicit symmetry-breaking parameter. 
In Sec.~\ref{sec4} and Sec.~\ref{chiral}, we generalize to the non-Abelian ${\rm SU}(N)$ and the chiral symmetry of QCD, ${\rm SU}(N)_{\rm L}\times {\rm SU}(N)_{\rm R}$, respectively. In Sec.~\ref{kubosection}, we derive the Kubo formula for the kinetic coefficients introduced in this work, and we conclude in Sec.~\ref{conclude}.

\section{Effective action approach to hydrodynamics}
\label{sec:EFT-review}

In this section, we briefly review the recently developed effective action approach to dissipative hydrodynamics, focusing on the case of internal globals symmetries without symmetry-breaking terms.
Our discussion will be pragmatic, and we refer the readers to Ref.~\cite{Glorioso:2018wxw} for more conceptual details. We will discuss a non-Abelian unitary symmetry $G$ in full generality, which can easily be applied to both ${\rm U}(1)$ and ${\rm SU}(N)_{\rm L} \times {\rm SU}(N)_{\rm R}$ as special cases.

We define dynamical variables of the nonequilibrium EFT as unitary group fields, $U_{1}(x)$ and $U_{2}(x)\in G$, defined on the two (forward and backward) SK contours, labeled by indices $1$ and $2$ respectively. 
We follow the convention in Ref.~\cite{Glorioso:2020loc} that the symmetry transformations $V_1(x)$ and $V_2(x)\in G$ on both contours, which are elevated to local transformations, act on these fields to the right; 
\be
U\to UV^{-1}=UV^{\dagger}\,,
\ee
where we omitted the contour labels for simplicity of presentation.

Moreover, we introduced the external gauge fields $A_{\mu 1}$ and $A_{\mu 2}$ on the two contours, which we will use to find the symmetry currents upon variation of the action with respect to them.
Their transformation rule is given by
\be
A_\mu\to VA_\mu V^{\dagger}+{\rm i}V\partial_\mu V^{\dagger}\,.
\ee
The useful gauge invariant combination is \be
B_\mu={\rm i} UD_\mu U^{\dagger}=UA_\mu U^{\dagger}+{\rm i} U\partial_\mu U^{\dagger},\ee
where we defined the covariant derivative as
\be
D_\mu U^{\dagger}=(\partial_\mu-{\rm i} A_\mu)U^{\dagger}\,.
\ee

In constructing the effective action for dissipative hydrodynamics based on underlying quantum theory, we take the classical limit $\hbar\to 0$ in the SK formalism.
In this limit, we assume that the variables in the two contours differ by $\cal O(\hbar)$, and the time scale of interests is much longer than the microscopic time scale of quantum fluctuations of typical energy $k_BT$, i.e., $\tau_{\rm q}=\hbar/(k_BT)$ (we set the Boltzman constant $k_B=1$ in the following).
Expanding the full effective action in the SK contours up to linear order in $\hbar$, and taking $\hbar\to 0$ limit, the path integral measure reduces to a classical action for hydrodynamics,
\be
\lim_{\hbar\to 0}{{\rm i} \over\hbar}S_{\rm SK}\to {\rm i} S_{\rm hydro}\,,
\ee
where $S_{\rm SK}$ is a microscopic action on the SK path and $S_{\rm hydro}$ is the effective action of dissipative hydrodynamics.
As we will see later, $S_{\rm hydro}$ should inherit the classical version of the original KMS symmetry that $S_{\rm SK}$ must possess.

Following Ref.~\cite{Glorioso:2020loc}, we introduce the variables in the classical action, $(U_{\rm r},\phi_{\rm a})$, as
\be
U_1=U_{\rm r}{\rm e}^{{{\rm i}\hbar\over 2}\phi_{\rm a}}=U_{\rm r}\left(1+{{\rm i}\hbar\over 2}\phi_{\rm a}\right)+\cdots \,,\quad U_2=U_{\rm r}{\rm e}^{-{{\rm i}\hbar\over 2}\phi_{\rm a}}=U_{\rm r}\left(1-{{\rm i}\hbar\over 2}\phi_{\rm a}\right)+\cdots\,,
\ee
where the elipsis denotes the higher order terms in $\hbar$. 
There is an ordering ambiguity between the ``classical value" $U_{\rm r}$ (or ``r-field") and the ``quantum fluctuation" variable $\phi_{\rm a}$ (or ``a-field"), while it simply redefines the latter by the unitary transformation with $U_{\rm r}$. This choice of definition for $\phi_{\rm a}$ gives rise to its many simple properties under symmetry transformations. 
Although the external gauge fields are arbitrary and do not need to follow the above rules of $\hbar$ expansion, we simply define
\be
A_{1}=A_{\rm r} + {\hbar\over 2}A_{\rm a}\,, \quad A_{2}=A_{\rm r}- {\hbar\over 2}A_{\rm a}\,;
\ee
otherwise, the $\hbar\to 0$ limit of the SK action diverges.

This assumption on the external gauge fields then requires us to consider the gauge transformations only of the following type ($v_{\rm a}^\dagger = v_{\rm a}$),
\be
\label{eq:V_12toar}
V_{1}=V_{\rm r}{\rm e}^{{{\rm i}\hbar\over 2}v_{\rm a}}=V_{\rm r}\left(1+{{\rm i}\hbar\over 2}v_{\rm a} \right)+\cdots\,, \quad V_{2}=V_{\rm r}{\rm e}^{-{{\rm i}\hbar\over 2}v_{\rm a}}=V_{\rm r}\left(1-{{\rm i}\hbar\over 2}v_{\rm a}\right)+\cdots \,,
\ee
under which the hydrodynamic fields and gauge fields transform as follows:
\be
 \begin{cases}
  U_{\rm r}\to U_{\rm r} V_{\rm r}^{\dagger}\,, \\
  \phi_{\rm a}\to V_{\rm r}(\phi_{\rm a}-v_{\rm a})V_{\rm r}^{\dagger}\,,
 \end{cases}
 \mathrm{and}\quad
 \begin{cases}
  A_{{\rm r}\mu}\to V_{\rm r}A_{{\rm r}\mu} V_{\rm r}^{\dagger}+{\rm i}V_{\rm r}\partial_\mu V_{\rm r}^{\dagger}\,,
  \\
 A_{{\rm a}\mu}\to V_{\rm r}(A_{{\rm a}\mu}+\partial_\mu v_{\rm a}-{\rm i}[A_{{\rm r}\mu},v_{\rm a}])V_{\rm r}^{\dagger}\,.
 \end{cases}
\ee
\cout{
\be
U_{\rm r}\to U_{\rm r} V_{\rm r}^{\dagger}
\,,
\quad 
\phi_{\rm a}\to V_{\rm r}(\phi_{\rm a}-v_{\rm a})V_{\rm r}^{\dagger}\,,
\ee
and the gauge fields transform as
\be
A_{{\rm r}\mu}\to V_{\rm r}A_{{\rm r}\mu} V_{\rm r}^{\dagger}+{\rm i}V_{\rm r}\partial_\mu V_{\rm r}^{\dagger}
\,,\quad 
A_{{\rm a}\mu}\to V_{\rm r}(A_{{\rm a}\mu}+\partial_\mu v_{\rm a}-{\rm i}[A_{{\rm r}\mu},v_{\rm a}])V_{\rm r}^{\dagger}.
\ee
}
We also decompose the gauge invariant $B_\mu$ terms as
\be
B_{1\mu}=B_{{\rm r}\mu}+ {\hbar\over2}B_{{\rm a}\mu}\,,\quad B_{2\mu}=B_{{\rm r}\mu}- {\hbar\over2}B_{{\rm a}\mu}\,,
\ee
where the gauge invariant r- and a-type terms are given by
\be
\label{eq:Br,Ba}
B_{{\rm r}\mu}={\rm i} U_{\rm r}D_\mu U_{\rm r}^{\dagger}=U_{\rm r} A_{{\rm r}\mu}U_{\rm r}^{\dagger}+{\rm i} U_{\rm r}\partial_\mu U_{\rm r}^{\dagger}\,,\quad B_{{\rm a}\mu}=U_{\rm r}(D_\mu\phi_{\rm a}) U_{\rm r}^{\dagger}\,,
\ee
with the covariant derivative, 
\begin{align}
\label{eq:covariant-phi_a}
D_\mu\phi_{\rm a}\equiv\partial_\mu\phi_{\rm a}-{\rm i}[A_{{\rm r}\mu},\phi_{\rm a}]+A_{{\rm a}\mu}\,.
\end{align}

The symmetry current in the classical limit and its quantum correction are defined as $J^\mu_{\rm r}\equiv \left(J^\mu_1+J^{\mu}_2\right)/2$ and $J^\mu_{\rm a}\equiv \left(J^\mu_1-J^{\mu}_2\right)/{\hbar}$ with $J^\mu_{\rm 1,2} = \delta S_{\rm SK}/\delta (A_{1,2})_\mu
$ on the microscopic action $S_{\rm SK}$. It can be easily shown that these currents are obtained by taking the variation of the effective action $S_{\rm hydro}$ with respect to the a-type and the r-type gauge fields $A_{\rm a}$ and $A_{\rm r}$, respectively,
\begin{align}
\label{eq:current}
J^\mu_{\rm r} = {\delta S_{\rm hydro}\over \delta A_{{\rm a} \mu}} \,,\quad J^\mu_{\rm a} = {\delta S_{\rm hydro}\over \delta A_{{\rm r} \mu}}\,.
\end{align}
Note that the r-type current is obtained by taking the variation with respect to  the a-type gauge field and vise visa.

The group fields $U_{1,2}$ represent the state of the system with regard to the symmetry $G$, which can be parametrized by the NG fields on the SK contours.
Roughly speaking, they are ``angle variables," canonically conjugate to the symmetry charges, whose corresponding chemical potentials are given by time derivatives of these fields.
In the symmetry-restored phase, the origin in the angle variable should not matter because the order parameter measuring such ``angle" is vanishing, and only the time derivative, i.e., the charge, gives a physical variable. 
As the fluid cells of hydrodynamics are locally independent of each other, the effective action should not depend on local redefinitions of the origin in the group fields. This indicates that the description of the symmetry-restored phase with the group fields $U_{1,2}$ possesses a gauge redundancy, which is called shift symmetry;
\be
\label{eq:shift-U_12}
U_{1,2}(x)\to \Lambda(\bm x)U_{1,2}(x)\,,
\ee
where $\Lambda(\bm x)\in G$ depends arbitrarily on space $\bm x$, but not on time (otherwise, it would affect the time derivatives). This effectively makes the time derivatives of $U$ fields, i.e., the charges, the basic degrees of freedom of the effective action.
For the hydrodynamic variables $U_{\rm r}$ and $\phi_{\rm a}$, the shift symmetry takes the form
\be
U_{\rm r}\to \Lambda U_{\rm r}\,,\quad \phi_{\rm a}\to\phi_{\rm a}\,,
\ee
and thus, the $G$-invariant building blocks behaves as
\be
\quad B_{{\rm a}\mu}\to \Lambda B_{{\rm a}\mu}\Lambda^{\dagger}\,,\quad B_{{\rm r}\mu}\to \Lambda B_{{\rm r}\mu}\Lambda^{\dagger}+{\rm i}\delta_{\mu i}\Lambda\partial_i\Lambda^{\dagger}\,,\quad (i=1,2,3).
\ee

On the other hand, in a symmetry-broken phase where $G$-symmetry is spontaneously broken down to its subgroup $H\subset G$, the shift symmetry should be reduced to the unbroken part only; $\Lambda(\bm x)\in H$.
The effective action for hydrodynamics then is allowed to contain $H\subset G$-part of $U$ fields as independent variables, and then $G/H$ corresponds to the NG bosons for the SSB. 
A nice example of this is the chiral symmetry of QCD, where $G={\rm SU}(N)_{\rm L}\times {\rm SU}(N)_{\rm R}$ and $H={\rm SU}(N)_{\rm V}$, which we will discuss in detail in Section \ref{chiral}. We also mention by passing that the gauge invariant $B_{\rm r}$-field in the $H$ part has been historically called the Stueckelberg field in EFTs for SSB.

The last ingredient we review is the KMS symmetry~\cite{Crossley:2015,Glorioso:2017fpd,Glorioso:2018wxw}. The KMS symmetry is the discrete $\mathbb{Z}_2$ symmetry of the generating functional in the SK formalism, consisting of imaginary-time translation with the inverse temperature $\beta=T^{-1}$ and the time-reversal transformation.
It acts on the external gauge fields as the following transformations:
\begin{subequations}
\label{eq:A-KMS-12}
\begin{align}
 A_{1\mu}(t,\bm x)
 &\to (-1)^\mu A^*_{1\mu}(-t,\bm x)=(-1)^\mu A^{\rm T}_{1\mu}(-t,\bm x)\,,\quad 
 \\
 A_{2\mu}(t,\bm x)
 &\to (-1)^\mu A^*_{2\mu}(-t-{\rm i}\hbar\beta,\bm x)\,,
\end{align}
\end{subequations}
where $(-1)^\mu=+1$ for $\mu=t$ and $-1$ for $\mu=i$, and we used the fact that the $A$ fields are hermitian matrices, i.e., $A^*=A^{\rm T}$.
We note that the complex conjugate operation originates from the time-reversal transformation.%
\footnote{\label{fn-2}The complex conjugate operation comes from the time reversal invariance of the microscopic theory under $A_\mu(t)\to (-1)^\mu A_\mu^*(-t)=(-1)^\mu A_\mu^{\rm T}(-t)$, which can be explicitly seen in several examples, e.g., Dirac fermions and complex scalar fields. 
One can also find this transformation rule by noting that the currents transform under time reversal as $j_\mu(t)\to(-1)^\mu j^*_\mu(-t)=(-1)^\mu j^{\rm T}_\mu(-t)$. 
Here, $*$ is a complex (not hermitian) conjugation acting on the hermitian matrix fields. The complex conjugate operation was missed in Ref.~\cite{Glorioso:2020loc}.
} 
The transpose operation does not matter in the ${\rm U}(1)$ case, but we keep them for the general ${\rm SU}(N)$ case. 
From now on, we shall omit the spatial argument of functions under the time-reversal operation. 
Considering that the external field appears in the covariant derivative of $U$ fields, we require the KMS symmetry to act on the dynamical fields $U$ as
\be
\label{eq:U-KMS-12}
U_1(t)\to U_1^*(-t)\,, \quad U_2(t)\to  U_2^*(-t-{\rm i}\hbar\beta)\,,
\ee
which enables us to respect the KMS symmetry manifestly in the low-energy EFT.

In the classical limit, the transformation rules of the dynamical and external fields in Eqs.~(\ref{eq:A-KMS-12}) and (\ref{eq:U-KMS-12}) become%
\footnote{Notation: $(\partial_t f)(-t) = f'(-t)$, where $f'(t)=\partial_t f(t).$}
\begin{align}
 A_{{\rm r}\mu}(t)
 &\to (-1)^\mu A^*_{{\rm r}\mu}(-t)\,,\quad A_{{\rm a}\mu}(t)\to (-1)^\mu\left[A^*_{{\rm a}\mu}(-t)+{\rm i}\beta (\partial_t A^*_{{\rm r}\mu})(-t)\right]\,,
 \\
 \label{eq:U,phi_a-KMS}
 U_{\rm r}(t)
 &\to U_{\rm r}^*(-t)\,,
 \hspace{54pt}
 \phi_{\rm a}(t)\to -\phi_{\rm a}^*(-t)+\beta (U_{\rm r}^{\rm T}\partial_t U_{\rm r}^*)(-t)\,,
\end{align}
\cout{
\be
A_{{\rm r}\mu}(t)\to (-1)^\mu A^*_{r\mu}(-t)\,,\quad A_{{\rm a}\mu}(t)\to (-1)^\mu\left[A^*_{a\mu}(-t)+{\rm i}\beta (\partial_t A^*_{r\mu})(-t)\right]\,,
\ee
and
\be
\label{eq:U,phi_a-KMS}
U_{\rm r}(t)\to U_{\rm r}^*(-t)\,,\quad \phi_{\rm a}(t)\to -\phi_{\rm a}^*(-t)+\beta (U_{\rm r}^{\rm T}\partial_t U_{\rm r}^*)(-t)\,,
\ee
}
respectively. Note that these transformations do not preserve the hermitian nature of $A_{{\rm a}\mu}$ and $\phi_{\rm a}$ fields since the second term comes from the imaginary time ${\rm i}\beta\hbar$.
As a consequence, the KMS transformations for their complex conjugate are not obtained by taking a complex conjugate operation to the above transformations. Instead, following the definition carefully, e.g., 
\be
\phi_{\rm a}^*=\lim_{\hbar\to 0}{1\over -{\rm i}\hbar}U_{\rm r}^{\rm T} (U_1^*-U_2^*)\,,
\ee
we derive the KMS transformations for the complex conjugate fields:
\be
\label{eq:KMS_Aphi_star}
A_{{\rm a}\mu}^*(t)\to (-1)^\mu\left[A_{{\rm a}\mu}(-t)+{\rm i}\beta (\partial_t A_{{\rm r}\mu})(-t)\right]\,,\quad
\phi_{\rm a}^*(t)\to  -\phi_{\rm a}(-t)-\beta (U_{\rm r}^\dagger\partial_t U_{\rm r})(-t)\,.
\ee
From these, it can be checked that the gauge invariant $B$-fields transform nicely as
\begin{align}
 B_{{\rm r}\mu}(t)
 &\to (-1)^\mu B_{{\rm r}\mu}^*(-t)=(-1)^\mu B_{{\rm r}\mu}^{\rm T}(-t)\,,
 \\
 B_{{\rm a}\mu}(t)
 &\to (-1)^\mu\left[B^{\rm T}_{{\rm a}\mu}(-t)+{\rm i}\beta (\partial_t B^{\rm T}_{{\rm r}\mu})(-t)\right]\,,
\end{align}
\cout{
\be
B_{{\rm r}\mu}(t)\to (-1)^\mu B_{{\rm r}\mu}^*(-t)=(-1)^\mu B_{{\rm r}\mu}^{\rm T}(-t)\,,\quad B_{{\rm a}\mu}(t)\to (-1)^\mu\left[B^{\rm T}_{a\mu}(-t)+{\rm i}\beta (\partial_t B^{\rm T}_{r\mu})(-t)\right]\,,
\ee
}
which is ultimately a consequence of the fact that the gauge transformation and the KMS transformation commute.

The effective action respecting all these symmetries is constructed in the gradient expansion of the dynamical variables $(U_{\rm r},\phi_{\rm a})$. 
It is also required from the unitarity of the microscopic theory that the action satisfies the condition, $S_{\rm hydro}[\Phi_{\rm a}\to-\Phi_{\rm a}]=-S_{\rm hydro}^*[\Phi_{\rm a}]$, where $\Phi_{\rm a}$ generically denotes ``a"-type fields. Specifically, terms containing odd numbers of ``a"-type fields should be real, and terms with even numbers of ``a"-type fields should be purely imaginary. Finally, a stability condition requires ${\rm Im}[S_{\rm hydro}]\ge 0$.

\section{A warm-up with approximate ${\rm U}(1)$ symmetry}
\label{sec:U(1)-warm-up}

Our main goal is to construct the effective action in a power series of a set of explicit symmetry-breaking parameters, $M$, which is responsible for the explicit symmetry breaking in the microscopic theory. 
Having in mind its application to QCD where $M$ is the quark mass matrix, we simplify our discussion by assuming that $M$ forms a simple multiplet under spurious symmetry transformations. 
In this section, we consider the simplest case of ${\rm U}(1)$ symmetry with a complex symmetry-breaking parameter $M$, which is promoted to a spurious complex field of charge $-1$. 

\subsection{Exact symmetry}

At zeroth order in $M$, i.e., the theory with the exact ${\rm U}(1)$ symmetry, the effective Lagrangian up to the second order in the derivative expansion is known to be \cite{Delacretaz:2021qqu},
\be
\label{eq:L0-U(1)}
\mathcal{L}_{\mathrm{U(1)}}^{(2)} =
\chi B_{{\rm a}t}B_{{\rm r}t}-f_\pi^2 B_{{\rm a}i}B_{{\rm r}i}-\chi^2\kappa B_{{\rm a}t}\left(\partial_t B_{{\rm r}t}-{\rm i}\beta^{-1}B_{{\rm a}t}\right)-\lambda B_{{\rm a}i}\left(\partial_t B_{{\rm r}i}-{\rm i}\beta^{-1}B_{{\rm a}i}\right)\,,
\ee
where $\chi$ is the linearized charge susceptibility,\footnote{{It is straightforward to generalize it to the case where $\chi$ is replaced with a general function of $B_{{\rm r}t}$}.} $f_\pi$ the NG decay constant, $\lambda$ the conductivity, and $\kappa$ a dissipative kinetic coefficient that plays an important role when the symmetry is spontaneously broken. Writing the ${\rm U}(1)$ field explicitly as $U_{\rm r}={\rm e}^{{\rm i}\phi_{\rm r}}$, we have the simple expressions for $B$-fields (\ref{eq:Br,Ba}) as
\be
\label{eq:U(1)-action-exat-sym}
B_{{\rm r}\mu}=\partial_\mu\phi_{\rm r}+A_{{\rm r}\mu}=D_\mu\phi_{\rm r}\,,\quad B_{{\rm a}\mu}=\partial_\mu\phi_{\rm a}+A_{{\rm a}\mu}=D_\mu\phi_{\rm a}\,.
\ee

Let us first consider the symmetry-restored phase where there is no SSB. 
The shift symmetry forbids the second term, i.e., $f_\pi=0$, and we see that the action can be written solely in terms of the chemical potential $\mu \equiv B_{{\rm r}t}$ without $\phi_{\rm r}$ explicitly appearing as we emphasized around Eq.~(\ref{eq:shift-U_12}). The action (\ref{eq:U(1)-action-exat-sym}) becomes the theory of diffusion \cite{Glorioso:2018wxw,Chen-Lin:2018kfl}. 
Note that the $\kappa$ term becomes a higher order term for the counting $\partial_t\sim \bm\nabla^2$ and can be neglected in the present case of the symmetry restored phase. Nevertheless, we shall keep this term for later purposes in describing the symmetry-broken phase.

The equation of motion (EOM) for $\phi_{\rm a}$ field reduces to the ordinary stochastic diffusion equation accompanied by the noisy constitutive relation for the charge current 
(using $\partial_tB_{{\rm r}i}=\partial_tD_{\rm i}\phi_{\rm r}=\partial_i D_t\phi_{\rm r}-E_i=\partial_iB_{{\rm r}t}-E_i$),
\be
\partial_\mu J_{\rm r}^\mu=0\,,\quad J^t_{\rm r}=\chi\mu-\chi^2\kappa\partial_t\mu+\xi_\kappa\,, \quad J_{\rm r}^i=-\lambda(\partial_i\mu-E_i)+\xi^i\,,
\ee
where $A_{{\rm a}\mu}$ is switched off after computing $J^\mu_{\rm r}$. 
We have changed the variable $\phi_{\rm a}$ involved in the expression followed by its original definition (\ref{eq:current}) into the noise terms $\xi_\kappa$ and $\xi^i$ with the noise correlations%
\footnote{
See, e.g., Ref.~\cite{tauber2014critical} for details of changing variables. 
One can equvalently understand this by identifying Eq.~\eqref{eq:L0-U(1)} as the effective action in the path-integral formalism of stochastic charge diffusion\cite{Martin:1973zz,Janssen:1976qag,DeDominicis1978}.
}
:
\be
\label{eq:noises-U(1)}
 \langle \xi_\kappa(x)\xi_\kappa(x')\rangle=2\chi^2\kappa \beta^{-1}\delta^{(4)}(x-x')\,,
 \quad
 \langle \xi^i (x) \xi^j (x') \rangle=2\lambda \beta^{-1}\delta^{ij} \delta^{(4)} (x-x')\,.
\ee
The noise correlators have been determined by $\propto \phi_{\rm a}^2$ terms in the action (\ref{eq:L0-U(1)}), or equivalently the coefficients in front of the $\propto \phi_{\rm a}$ terms in the original expression of the current. 
The changing variable enables us to compute certain correlation functions in the later sections without discussing the possible contact terms in the Schwinger-Dyson equations for the (r,a) variables.

On the other hand, the absence of the shift symmetry in the symmetry-broken phase allows $f_\pi\neq 0$, and thus, $\phi_{\rm r}$ becomes a propagating gapless mode of linear dispersion relation with a finite velocity,
\begin{align}
\label{eq:vs_m=0}
v_{\rm s}^2=\frac{f_\pi^2}{\chi}\,,
\end{align}
which is the NG boson associated with spontaneous ${\rm U}(1)$ symmetry breaking. 
Note that the expression for (\ref{eq:vs_m=0}) is valid only in the exact symmetry limit described by the effective Lagrangian~(\ref{eq:L0-U(1)}); we will discuss its correction from the symmetry-breaking parameter later. 

The equation of motion (EOM) yields the conservation equation $\partial_\mu J^\mu_{\rm r}=0$ with
\be
\label{eq:curret-expression}
J^t_{\rm r}=\chi D_t\phi_{\rm r}-\chi^2\kappa \partial_tD_t\phi_{\rm r}+\xi_\kappa\,,\quad J_{\rm  r}^i=-f_\pi^2D_{\rm i}\phi_{\rm r}-\lambda(\partial_i \mu-E_i)+\xi^i\,.
\ee
where $\xi_\kappa$ and $\xi^i$ are the same thermal noises given in Eq.~\eqref{eq:noises-U(1)}. 
Equivalently, we can also rewrite these in two coupled equations of $(\phi_{\rm r},\,\mu)$,
\be
D_t\phi_{\rm r}=\mu\,,\quad \chi\partial_t\mu=f_\pi^2\partial_i D_{\rm i}\phi_{\rm r} +\lambda\partial_i(\partial_i\mu-E_i)+\chi^2\kappa\partial_t^2\mu-\partial_t\xi_\kappa-\partial_i\xi^i\,,
\ee
where the first equation is called the Josephson relation. Other ways of expressing the same equations exist, truncating at the second order in derivatives and using the lowest order EOM, $\chi\partial_tD_t\phi_{\rm r}=f_\pi^2\partial_iD_{\rm i}\phi_{\rm r}$. For example, we can move the dissipative effect from $\kappa$ to the Josephson relation, by defining $\tilde\mu\equiv J^t_{\rm r}/\chi=D_t\phi_{\rm r}-\chi\kappa\partial_tD_t\phi_{\rm r}+\xi_\kappa/\chi=D_t\phi_{\rm r}-f_\pi^2\kappa\partial_iD_{\rm i}\phi_{\rm r}+\xi_\kappa/\chi$ and $\tilde\xi_\kappa\equiv\xi_\kappa/\chi$, which results in 
\be
D_t\phi_{\rm r}=\tilde\mu+f_\pi^2\kappa \partial_iD_{\rm i}\phi_{\rm r} -\tilde\xi_\kappa\,, \quad \chi\partial_t\tilde\mu=f_\pi^2\partial_iD_{\rm i}\phi_{\rm r}+\lambda\partial_i(\partial_i\tilde\mu-E_i)-\partial_i\xi^i\,,
\ee
where $\langle \tilde\xi_\kappa (x) \tilde\xi_\kappa (x')\rangle=2\kappa T \delta^{(4)} (x-x')$.
Both $\kappa$ and $\lambda$ contribute to the leading diffusive term in the following dispersion relation for the frequency $\omega$ and momentum $\bm{k}$:
\be
 \omega (k) = \pm v_s k- \frac{{\rm i}}{2} 
 \left( \frac{\lambda}{\chi} + f_\pi^2 \kappa \right) k^2
 + {\cal O}(k^3),
\ee
with $k\equiv |\bm{k}|$ and the propagating velocity $v_s$ already defined in Eq.~\eqref{eq:vs_m=0}.

\subsection{Approximate symmetry}

We next consider effects from the explicit symmetry-breaking terms containing the symmetry-breaking parameter $M$.
Although we will eventually set $M_1$ and $M_2$ on the two SK contours to their physical value as $M_1=M_2=M = \mathrm{const.}$, we will initially treat them as spurious fields that transform under the doubled symmetry. 
In other words, in constructing the effective action, we consider each of $M_1$ and $M_2$ ($M_1\neq M_2$ in general) as independent spurious fields (functions of space and time), on which the doubled ${\rm U} (1)$ symmetries act.
One sees that this gives an extended notion of the background field in the SK formalism, and thus, $M_1$ and $M_2$ are not dynamical fields. 
Indeed, once
the action is constructed respecting the full spurious symmetry, we replace $M_{1,2}$ with their physical value.

We then assume that the microscopic theory remains invariant by assigning a spurious transformation rule to the $M$ fields as
\be
M_1\to M_1 V_1^{\dagger}\,,\quad M_2\to M_2 V_2^{\dagger}\,.
\ee
with $V\in {\rm SU}(N)$ for latter purpose. 
Expressing
\be 
M_{1}=M_{\rm r}\left(1+{{\rm i}\hbar\over 2}m_{\rm a}\right)+\cdots \,,\quad M_{2}=M_{\rm r}\left(1-{{\rm i}\hbar\over 2}m_{\rm a}\right)+\cdots \,,
\ee
we have the gauge transformations as [recall Eq.~(\ref{eq:V_12toar})]
\be
\label{eq:M-sperious-gauge-trans}
M_{\rm r}\to M_{\rm r}V_{\rm r}^{\dagger} ,\quad m_{\rm a}\to V_{\rm r}(m_{\rm a}-v_{\rm a})V_{\rm r}^{\dagger}\,,
\ee
where it simplifies $m_{\rm a} \rightarrow m_{\rm a}-v_{\rm a}$ for the ${\rm U}(1)$ case.

The KMS symmetry transformation should also act on the $M$ fields since the theory would be time reversal invariant if the theory without $M$ fields is time reversal invariant. 
The $M$ fields also transform according to the rules of time reversal transformations. 
In the phase where the ${\rm U}(1)$ symmetry is spontaneously broken, we also assume that the time reversal symmetry is not spontaneously broken by the order parameter (see Ref.~\cite{Hongo:2018ant} for the discussion about the KMS symmetry in time-translational symmetry breaking case). 
Therefore, we require the action to be KMS invariant. 
We assume that the time reversal parity of the $M$ fields is $+1$, i.e., the KMS transformation is the same as that for the dynamical $U$ fields [see Eq.~(\ref{eq:U,phi_a-KMS})];%
\footnote{That is, under the time reversal, we have $M(t)\to M^*(-t)$.
This remains the same even in non-Abelian case discussed later, where $M$ is matrix-valued.}
\be
\label{eq:KMS-Mrma}
M_{\rm r}(t)\to M_{\rm r}^*(-t)\,,\quad m_{\rm a}(t)\to -m_{\rm a}^*(-t) +\beta (M_{\rm r}^*)^{-1}(\partial_t M_{\rm r}^*)(-t)\,,
\ee
and (note the sign flip in the second term as before for $\phi_{\rm a}^*$),
\be
\label{eq:KMS-mastar}
m_{\rm a}^*(t)\to -m_{\rm a}(-t) -\beta M_{\rm r}^{-1}(\partial_t M_{\rm r})(-t)\,.
\ee
Finally, the shift symmetry does not act on the $M$ fields since it refers to the emergent gauge redundancy of our description of hydrodynamics rather than the symmetry of the microscopic theory.
Since we will replace $M$ fields with their physical value at the end, i.e., $M_{\rm r}=M$, a constant, and $m_{\rm a}=0$, we will organize the action in the power series of $M_{\rm r}$.

\subsubsection{The first order in $M$}

At the first order in $M_{\rm r}$, there is no action that respects all the identified symmetries, including the shift symmetry. 
Therefore, if there is no ${\rm U}(1)$ SSB, the effect from the explicit symmetry breaking starts to appear only in the second order of $M_{\rm r}$, which we will discuss in Sec.~\ref{sec:mass-second-order}. 

We can prove the above statement as follows; write the first-order term in the effective action as $M_{\rm r} \cdot f(U_{\rm r},D_\mu U_{\rm r},\phi_{\rm a},m_{\rm a},\cdots)$.
Since $M_{\rm r}$ has charge $-1$ under the gauge symmetry $V_{\rm r}$, the gauge invariance requires that $f$-term as a whole has charge $+1$.
Consider a homogeneous transformation of shift symmetry $\Lambda$. Due to the Abelian nature of ${\rm U}(1)$, its action to the fields is found to be identical to the ${\rm U}(1)$ symmetry transformation with $V_{\rm r}^{-1}=\Lambda$ and $v_{\rm a}=0$. Therefore, we see that $f$-term has charge $-1$ under the shift symmetry, and transforms as $f\to f \Lambda^{-1}$. 
Since $M_{\rm r}$ does not transform under the shift symmetry, the term in consideration is inconsitent with the shift symmetry.
Note that the convariant derivatives of $M_{\rm r}$, e.g., $D_\mu M_{\rm r}=\partial_\mu M_{\rm r}+{\rm i}M_{\rm r}A_{{\rm r}\mu}$ share the same charge as $M_{\rm r}$, and are allowed to appear in the action. Our result equally applies to them as well.

On the other hand, the shift symmetry is absent in the ${\rm U}(1)$ SSB phase.
Thus, we can find the leading terms at $\mathcal{O} (M)$ respecting all other symmetries as
\be
\label{eq:L_mass-linear}
{\cal L}_M= {\sigma}(M_1U_1^{*}-M_2U_2^{*})/\hbar +{\rm c.c.}=-{\rm i}\sigma \mathcal{M}_{\rm a} +{\rm c.c.}\,,
\ee
with a real constant $\sigma$. 
Here, for later purposes, we introduced a gauge invariant (shift covariant) quantity linear both in $M_{\rm r}$ and a-fields: 
\begin{align}
\label{eq:calMa}
\mathcal M_{\rm a} \equiv M_{\rm r}(\phi_{\rm a}-m_{\rm a})U_{\rm r}^{\dagger}\,.
\end{align}
We also note that, for the present U(1) case, $U_{\rm r}^{\dagger}$ is replaced by $U_{\rm r}^{*}$.
Moreover, the KMS invariance is respected up to the total derivative term as%
\footnote{
In fact, the KMS transformation acts as $\sigma\to \sigma^*$. 
The constant $\sigma$ is the order parameter of the SSB, and the reality condition of $\sigma$ is essentially our assumption that the time reversal symmetry is not spontaneously broken by the order parameter.
}
\be
{\cal L}_M \rightarrow \mathcal{L}_M +{{\rm i}\sigma\beta}\partial_t\left[M_{\rm r}^* U_{\rm r} (-t) + M_{\rm r} U_{\rm r}^* (-t)\right] \,.
\ee
We will assume that the physical value of $M_{\rm r}=M$ 
is real and positive, which is the stability condition for $\phi_{\rm r}$ field, and we can always redefine the origin of $U_{\rm r}$ to satisfy this condition.
As we will shortly see, this term makes the NG boson $\phi_{\rm r}$ massive with $m_{\rm NG}^2=2\sigma M/\chi$; the analogy to what happens in QCD is clear:
Indeed, Eq.~(\ref{eq:L_mass-linear}) corresponds to the leading order term linear in $M$ in the $\mathcal O(p^2)$ chiral Lagrangian \cite{Weinberg:1978kz,Gasser:1983yg,Gasser:1984gg,Georgi:1984zwz,Scherer-Schindler2011} on the SK contour.

Let us now investigate some consequences of the $\mathcal{O} (M)$ term by adding Eq.~(\ref{eq:L_mass-linear}) to the symmetric one~\eqref{eq:L0-U(1)}.
Firstly, the linearized EOM folowing from $\mathcal{L}_{\mathrm{U(1)}}^{(2)} + {\cal L}_M$ can be expressed up to second order in gradients as
\begin{subequations}
\label{eq:eom-damp-U(1)}
\begin{align}
D_t\phi_{\rm r}&=\tilde\mu+
\chi\kappa ( v_{\rm s}^2\partial_iD_{\rm i}\phi_{\rm r}-m_{\rm NG}^2\phi_{\rm r}) -\tilde\xi_\kappa\,, \\
\chi\partial_t\tilde\mu&=\chi(v_{\rm s}^2\partial_iD_{\rm i}\phi_{\rm r}-m_{\rm NG}^2\phi_{\rm r})+\lambda\partial_i(\partial_i\tilde\mu-E_i)-\partial_i\xi_i\,,
\end{align}
\end{subequations}
with the noise correlations for $(\tilde\xi_\kappa\,,\ \xi^i)$ unchanged as before. 
In writing this, we used the lowest order EOM, $\chi\partial_tD_t\phi_{\rm r}=f_\pi^2\partial_iD_{\rm i}\phi_{\rm r}-\chi m_{\rm NG}^2\phi_{\rm r}$, and $\tilde\mu\equiv J^t_{\rm r}/\chi$.
This form of the EOM was obtained previously, using the second law of thermodynamics \cite{Grossi:2020ezz}, the Poisson bracket method with dissipation \cite{Nishimura:2023czx}, or the locality of the response function to an external source for $\phi_{\rm r}$ field \cite{Delacretaz:2021qqu}.

By solving the linearized EOM \eqref{eq:eom-damp-U(1)}, we obtain the following dispersion relation:
\be
\label{eq:dispersion-U(1)}
\omega (k) =\pm\sqrt{v_{\rm s}^2 k^2+m_{\rm NG}^2}-{{\rm i}\over 2}\left[\frac{\lambda}{\chi} k^2+\chi \kappa(v_{\rm s}^2k^2+m_{\rm NG}^2)\right],
\ee
where we have kept the relativistic form of the real part (first term) with $m_{\rm NG}^2=2\sigma M/\chi$ (It is worth noticing the counting $m_{\rm NG}^2 = \mathcal{O} (M)$ for later purpose). 
We also note that the combination $(v_{\rm s}^2\partial_iD_{\rm i}\phi_{\rm r}-m_{\rm NG}^2\phi_{\rm r})$ appearing in the EOM \eqref{eq:eom-damp-U(1)} captures a feature in the dispersion relation, i.e., the effect of $m_{\rm NG}^2$ always appears via the replacement $v_{\rm s}^2 k^2\to v_{\rm s}^2k^2+m_{\rm NG}^2$.
As pointed out in Ref.~\cite{Son:2002ci}, the term in Eq.~(\ref{eq:L_mass-linear}) also contributes to the damping of the pseudo-NG boson in a definite way.

The nonvanishing damping term in $\bm k\to 0$ limit describes an exponential decay of the total ${\rm U}(1)$ charge in the system. 
This is the leading mechanism of ${\rm U}(1)$ charge relaxation of the system due to an explicit symmetry breaking, which is of the first order in the symmetry breaking parameter $M$ (or $m_{\rm NG}^2$). As the damping rate at $\bm k=0$ is proportional to $\chi \kappa m_{\rm NG}^2$ (i.e., $\kappa\sigma M$), it is an effect that arises from the interplay between the kinetic coefficient $\kappa$ and both the spontaneous and the explicit symmetry breakings, $\sigma$ and $M$. Physically, this damping occurs because the charge in the symmetry-broken phase is carried by the pseudo-NG boson, which decays at finite $m_{\rm NG}$.

To illustrate the effect of the explicit symmetry breaking, it is useful to investigate the current conservation law and its explicit violation by $M$.
The ${\rm U}(1)$ current, which is obtained by taking variation with respect to gauge fields, $J^\mu_{\rm r}=\delta S/\delta A_{{\rm a}\mu}$, is not modified by ${\cal L}_M$ as it does not contain gauge fields.
The EOM can be rephrased as the conservation law, which is modified at first order in the explicit symmetry breaking $M_{\rm r}$,
\be
\partial _\mu J^{\mu}_r={\rm i}\sigma M_{\rm r}(U_{\rm r}-U_{\rm r}^*)\simeq -\chi m_{\rm NG}^2 \phi_{\rm r}\,, 
\ee
where we linearized in $\phi_{\rm r}$ in the last expression. 
We note the nonvanishing right-hand side corresponds to the term appearing in the PCAC relation in QCD. As we have shown, this mechanism or the right-hand side of the above equation is absent in the phase where there is no SSB, and the leading mechanism for charge relaxation in that phase is of the second order in $M$, as we will show in Sec.~\ref{sec:mass-second-order}.

When we go to the next order of derivative expansion while keeping the mass expansion to the first order, we find two additional terms that are KMS invariant up to the total derivative,
\begin{align}
\label{eq:cross-Lag}
\ell_{\rm t}\left[{\rm i} B_{\rm at} D_t (M_{\rm r}U_{\rm r}^*) -(D_t B_{\rm rt})\mathcal{M}_{\rm a}\right] - \ell_{\rm s} \left[{\rm i} B_{\rm ai}D_i(M_{\rm r}U_{\rm r}^*)-(D_i B_{\rm ri})\mathcal{M}_{\rm a}\right] + {\rm c.c.}\,,
\end{align}
which is real and thus contributes to non-dissipative effects. In fact, we find corresponding terms in the next to leading order $\mathcal O(p^4)$ chiral Lagrangian in vacuum \cite{Gasser:1984gg}, which reduces to
\begin{align}
\label{eq:L4-chiral-Lag}
(D_\mu U D^\mu U^* ) MU^* + {\rm c.c.} = {\rm i}  B^\mu D_\mu (MU^*) + {\rm c.c.}
\end{align}
for the U(1) symmetry case.
One can show that (\ref{eq:L4-chiral-Lag}) is equivalent to (\ref{eq:cross-Lag}) when it is put on the SK contour. Note here that the Lorenz symmetry of (\ref{eq:L4-chiral-Lag}) is broken (to spatial rotational symmetry), which gives two independent parameters, $\ell_{\rm t}$ and $\ell_{\rm s}$, for the time and special components, respectively.

From Eq.~(\ref{eq:cross-Lag}), we have corrections to both the current (\ref{eq:curret-expression}) and the conservation law (\ref{eq:dJ-gamma}),
\begin{align}
\label{eq:cross-corection}
\Delta J_{\rm r}^t&={\rm i} \ell_{\rm t} D_t(M_{\rm r}U_{\rm r}^*-M_{\rm r}^*U_{\rm r})\,, \quad \Delta J_{\rm r}^i=-{\rm i} \ell_{\rm s} 
 D_i(M_{\rm r}U_{\rm r}^*-M_{\rm r}^*U_{\rm r}) \,, \\
\Delta(\partial_\mu J^\mu_{\rm r})&=-\ell_{\rm t} (D_t B_{\rm rt})(M_{\rm r}U_{\rm r}^* +M_{\rm r}^*U_{\rm r}) + \ell_{\rm s} (D_i B_{\rm ri})(M_{\rm r}U_{\rm r}^* +M_{\rm r}^*U_{\rm r}) \,,
\end{align}
respectively. These lead to non-dissipative corrections to the Josephson relation and the pion velocity linear in $M_{\rm r}$. 
As a result, the relation between the velocity, decay constant, and susceptibility in the chiral limit (\ref{eq:vs_m=0}) receives $\mathcal{O}(M)$ correction upon choosing the physical value $M_{\rm r}=M$,
\begin{align}
\label{eq:v-updated}
v_{\rm s}^2=\frac{f_\pi^2 + 4 M \ell_{\rm s}}{\chi + 4 M \ell_{\rm t} }\,.
\end{align}
On the other hand, (\ref{eq:cross-Lag}) does not affect the expression for the dispersion relation given by (\ref{eq:dispersion-U(1)}) as a function of $v_{\rm s}$.

Similar corrections to the velocity of the NG field have been first pointed out in Ref.~\cite{Armas:2021vku} within the thermodynamics approach based on the second law, and subsequently in Ref.~\cite{Delacretaz:2021qqu} within the nonequilibrium EFT (which called them ``pinning effects"). Our expression (\ref{eq:cross-Lag}) is fully non-linear in the NG field and is also invariant under the spurious symmetry transformation (\ref{eq:M-sperious-gauge-trans}). 
We find that these terms already exist in the chiral effective theory in vacuum as higher order gradient terms, and do not represent new hydrodynamics effects.

\subsubsection{The second order in $M$}
\label{sec:mass-second-order}
We next consider the effective action that are second order in $M_{\rm r}$.
A useful method to construct a KMS invariant term is to rely on the $\mathbb{Z}_2$-symmetry of the KMS transformation \cite{Glorioso:2018wxw}.

Under the KMS transformation, we have $\Phi(t)\to (K\Phi)(-t)$ for a generic field $\Phi$, which defines a map $K: \Phi(t)\to (K\Phi)(t)$ (note that this map leaves the argument $t$ the same).
For example, $KU=U^*$, $K\phi_{\rm a}=-\phi_{\rm a}^*+\beta (U_{\rm r}^{\rm T}\partial_t U_{\rm r}^*)$ and $K\phi_{\rm a}^*=-\phi_{\rm a}-\beta(U_{\rm r}^\dagger\partial_t U)$ as we have obtained in Eqs.~(\ref{eq:U,phi_a-KMS}) and (\ref{eq:KMS_Aphi_star}) (note the sign flip in the second term in $K\phi_{\rm a}$ and $K\phi_{\rm a}^*$), and similarly for $M$ fields.
From the definition of $K$, when $K$ acts on a time derivative, it changes the sign of the time derivative, i.e., $K(\partial_t \Phi)=-\partial_t(K\Phi)$.
It can be checked easily that $K^2=1$, i.e., $K$ is a $\mathbb{Z}_2$-action. 

Then, consider a term in the effective action, $F(t)(KF)(t)$, where $F$ is an arbitrary term constructed from the fields $\Phi$. Under the KMS transformation, this becomes $(KF)(-t)F(-t)=(KF)(t')F(t')$ where $t'=-t$, and the action is KMS invariant with the time direction reversed.
In fact, the previous dissipative terms in the action (\ref{eq:L0-U(1)}) can be written as
\be
-{\rm i}\lambda\beta^{-1} B_{{\rm a}i} (KB_{{\rm a}i})+{\rm i}\kappa\chi^2\beta^{-1} B_{{\rm a}t}(KB_{{\rm a}t})\,.
\ee

In constructing the KMS invariant terms with $M$ fields using the above method, it is crucial to note that the time derivative appearing in the KMS transformations for ``a"-type fields is the ordinary, not gauge covariant, derivative.
This means that the above method would result in an expression that is not gauge invariant in general unless $F$ is gauge invariant.
For the gauge invariant $F$, its ordinary time derivative (equal to the covariant derivative) is consistent with the gauge symmetry. 
This can also be understood from the fact that the gauge transformation and the KMS transformation commute.
Since we are interested in a dissipative term, we consider $F$ that is linear in ``a"-type field; then $F(KF)$ would contain a term quadratic in ``a"-type fields, which should then be purely imaginary due to the unitarity condition, and hence dissipative. Finally, a quadratic action term in $M_{\rm r}$ would require us to consider $F$ linear in $M_{\rm r}$ field.

The $F$ satisfying these requirements at zeroth order in the derivative is unique;
\be
F= c_1\mathcal{M}_{\rm a}+c_2\mathcal{M}_{\rm a}^*\,,
\ee
where $\mathcal{M}_{\rm a}$ is given by (\ref{eq:calMa}) with $c_{1,2}$ being constants. Using Eqs.~(\ref{eq:U,phi_a-KMS}) and (\ref{eq:KMS-Mrma}) with $U_{\rm r}^*U_{\rm r} =1$, we obtain
\begin{align}
K\mathcal{M}_{\rm a} &= - \mathcal{M}_{\rm a}^* -\beta \partial_t(M_{\rm r}^*U_{\rm r})\,,
\end{align}
and similarly using (\ref{eq:KMS_Aphi_star}) and (\ref{eq:KMS-mastar}), we obtain
\begin{align}
K\mathcal{M}_{\rm a}^* &= - \mathcal{M}_{\rm a} +\beta \partial_t(M_{\rm r}U_{\rm r}^{*}) \,.
\end{align}
Thus
\be
KF=-c_1\mathcal{M}_{\rm a}^*-c_2\mathcal{M}_{\rm a}-\beta\left[c_1\partial_t(M_{\rm r}^*U_{\rm r})-c_2\partial_t(M_{\rm r}U_{\rm r}^{*})\right]\,.
\ee
It can be shown that for arbitrary $c_{1,2}$, we have
\be
F (KF)={\rm i}(c_1^2+c_2^2)L_1+2{\rm i} c_1c_2 L_2-{\beta\over 2}(c_1^2-c_2^2)L_3\,.
\ee
where $L_{1,2,3}$ are the KMS invariant terms that satisfy the unitarity and stability conditions independently,
\begin{align}
L_1&={\rm i}\mathcal{M}_{\rm a} \mathcal{M}_{\rm a}^*+{{\rm i}\beta\over 2}\left[\mathcal{M}_{\rm a} \partial_t(M_{\rm r}^* U_{\rm r})-\mathcal{M}_{\rm a}^*\partial_t(M_{\rm r} U_{\rm r}^{*})\right]\,,\\
L_2&= {{\rm i}\over 2}\left[\mathcal{M}_{\rm a}^2+(\mathcal{M}_{\rm a}^*)^2\right]-{{\rm i}\beta\over 2}\left[\mathcal{M}_{\rm a}\partial_t(M_{\rm r}U_{\rm r}^{*})-\mathcal{M}_{\rm a}^*\partial_t(M_{\rm r}^*U_{\rm r})\right]\,,\\
L_3&= \mathcal{M}_{\rm a}\partial_t(M_{\rm r}^*U_{\rm r})+\mathcal{M}_{\rm a}^*\partial_t(M_{\rm r}U_{\rm r}^{*})\,.
\end{align}
Here, pure imaginaries in $L_{1,2}$ for later purposes. Therefore, a general KMS invariant action consistent with unitarity is an arbitrary linear combination of $L_{1,2,3}$ with positive real coefficients.
The $L_3$ is linear in ``a"-type field, and hence real and non-dissipative. When we use the physical value of $M_{\rm r}=M$, $m_{\rm a}=0$, and also $\phi_{\rm a}=\phi_{\rm a}^*$ for ${\rm U}(1)$, this term vanishes identically,
\be
|M|^2 \phi_{\rm a}\partial_t(U_{\rm r} U_{\rm r}^{*})=0\,,
\ee
where we used the ${\rm U}(1)$ property, $U_{\rm r} U_{\rm r}^{*}=1$. We will not further discuss about this term in the following.
On the other hand, the $L_{1,2}$ contains the imaginary valued terms quadratic in a-fields, and they are dissipative.

Under the shift symmetry under which $U$ has charge $+1$, $L_{1}$ is invariant, but $L_2$ is not. Therefore, in the case the ${\rm U}(1)$ symmetry is not spontaneously broken, we impose the invariance under the shift symmetry, $L_{2}$ is not allowed in the action, and the action contains the following terms in general, with a kinetic coefficient $\gamma_1$,
\be
\gamma_1\beta^{-1} L_1={{\rm i}\gamma_1\over 2}\left[\mathcal{M}_{\rm a} \partial_t(M_{\rm r}^* U_{\rm r})-\mathcal{M}_{\rm a}^*\partial_t(M_{\rm r} U_{\rm r}^{*})\right]+{\rm i}\gamma_1\beta^{-1}\mathcal{M}_{\rm a}\mathcal{M}_{\rm a}^*\,.
\ee
The first piece, which is linear in ``a"-type field, contributes to the classical EOM for the ${\rm U}(1)$ charge density and the second piece represents the corresponding noise according to the fluctuation-dissipation theorem. To understand the physics meaning of these terms, we first note that the ordinary time derivative is equal to the gauge covariant derivative for a gauge invariant term, so the first piece can be written as
\be
\gamma_1\beta^{-1} L_1 = {{\rm i}\gamma_1\over 2}\left[\mathcal{M}_{\rm a} D_t(M_{\rm r}^* U_{\rm r}^{\rm T})-\mathcal{M}_{\rm a}^*D_t(M_{\rm r} U_{\rm r}^*)\right]+{\rm i}\gamma_1\beta^{-1}\mathcal{M}_{\rm a}\mathcal{M}_{\rm a}^*\,.
\ee
Using the Leibniz rule for the covariant derivative and assuming that the covariant derivative of $M_{\rm r}$ vanishes,\footnote{This is a reasonable assumption based on the fact that a dynamical $M$-field, e.g., the Higgs field in the Standard Model, would adjust itself to satisfy this condition in a more fundamental theory. A non-vanishing $D_t M$ can act as an external source or sink of the symmetry charge, which we assume is absent.} we arrive at
\be
\label{eq:L_1_U(1)-final}
\gamma_1\beta^{-1} L_1 =-\gamma_1|M_{\rm r}|^2 \phi_{\rm a}B_{{\rm r}t} + {\rm i} \gamma_1 |M_{\rm r}|^2\beta^{-1} \phi_{\rm a}^2\,,
\ee
where have set $m_{\rm a}=0$ and and used the fact that $\phi_{\rm a}^*=\phi_{\rm a}$. Note also that $B_{{\rm r}t}={\rm i} U_{\rm r}D_t U_{\rm r}^*$ for ${\rm U}(1)$. The EOM from this can be written as a violation of the charge conservation law with stochastic noise variable $\xi$,
\be
\partial_\mu J_{\rm r}^\mu
=-\gamma_1|M_{\rm r}|^2 \mu + \xi\,,
\ee
where  
\be
\langle \xi(x)\xi(x')\rangle=2\gamma_1 \beta^{-1} |M_{\rm r}|^2\delta^{(4)}(x-x')\,.
\ee
We recalled the identification of the chemical potential $\mu=B_{{\rm r}t}$. This is a stochastic charge relaxation equation, which gives a linear relaxation of charges with a relaxation rate $\Gamma_1=\gamma_1|M_{\rm r}|^2/\chi=\gamma_1 M^2/\chi$ when $\bm k\to 0$. 
Note that $L_{1,2,3}$ terms do not modify the symmetry current, $J_{\rm r}^\mu$ computed by (\ref{eq:current}), since they do not contain the gauge fields. As (\ref{eq:L_1_U(1)-final}) is the leading order mechanism of charge relaxation in the phase where the symmetry is not spontaneously broken, we find that the charge relaxation in this phase is of second order in $M$.

In the SSB phase, both $L_1$ and $L_2$ are allowed, and the action contains in general,
\be
\gamma_1\beta^{-1} L_1+\gamma_2\beta^{-1} L_2\,.
\ee
The $L_2$ term is explicitly given by
\be
\gamma_2\beta^{-1}L_2=
-{\gamma_2\over 2}\left[(M_{\rm r}^* U_{\rm r})^2+(U_{\rm r}^* M_{\rm r})^2\right]\phi_{\rm a} B_{{\rm r}t}+{{\rm i}\over 2}
\gamma_2 \beta^{-1}\left[(M_{\rm r}^* U_{\rm r})^2+(U_{\rm r}^* M_{\rm r})^2\right]\phi_{\rm a}^2\,.
\ee
We note that the stability condition, i.e., the positivity of the imaginary part of the action, imposes a condition, $\gamma_1>|\gamma_2|$ for arbitrary $\phi_{\rm r}$ with $U={\rm e}^{{\rm i}\phi_{\rm r}}$ and the physical mass $M_{\rm r}=M$.  
Linearizing 
in $\phi_{\rm r}$, 
we find
\be
\gamma_2\beta^{-1}L_2=-\gamma_2M^2\phi_{\rm a} B_{{\rm r}t}+{\rm i}
\gamma_2 \beta^{-1}M^2 \phi_{\rm a}^2+{\cal O}(\phi_{\rm r}^2)\,,
\ee
which takes the same form to $L_1$ term, that is, the effect of $L_2$ is identical to that of $L_1$ for linearized dynamics of $\phi_{\rm r}$, although they are different at non-linear level.

The conservation law, or the EOM, becomes
\be
\label{eq:dJ-gamma}
\partial_\mu J_{\rm r}^\mu={\rm i}\sigma M_{\rm r}(U_{\rm r}-U_{\rm r}^*)-\gamma_1|M_{\rm r}|^2B_{{\rm r}t}-{\gamma_2\over 2}\left[(M_{\rm r}^* U_{\rm r})^2+(U_{\rm r}^* M_{\rm r})^2\right]B_{{\rm r}t}+\xi\,,
\ee
with the noise correlation
\begin{align}
\langle\xi(x)\xi(x')\rangle &= 2T\left\{\gamma_1|M_{\rm r}|^2+{\gamma_2\over 2} \left[(M_{\rm r}^* U_{\rm r})^2+(U_{\rm r}^* M_{\rm r})^2\right]\right\}\delta^{(4)}(x-x') \notag\\
&\quad \simeq 2\gamma TM^2\delta^{(4)}(x-x')\,,
\end{align}
where $\gamma \equiv \gamma_1 + \gamma_2$ and we have linearized in $\phi_{\rm r}$ field and used the physical value $M_{\rm r}=M$ in the last line. 
The linearized dispersion relation for $\phi_{\rm r}$, including the effects from $L_{1,2}$ is found to be
\be
\label{eq:dispersion-U(1)-SSB}
\omega (k) =\pm\sqrt{v_{\rm s}^2 k^2+m_{\rm NG}^2}-{{\rm i}\over 2}\left[\frac{\lambda k^2}{\chi}+\chi\kappa (v_{\rm s}^2 k^2+m_{\rm NG}^2)+ \frac{\chi\gamma m_{\rm NG}^4}{4\sigma^2} \right]\,.
\ee
The damping effect caused by $L_1$ and $L_2$, $\chi\gamma m_{\rm NG}^4/(4\sigma^2) = \gamma M^2/\chi$ is sub-leading compared to the same effect from $\chi \kappa m_{\rm NG}^2=\kappa\sigma M$, which is linear in $M$ as discussed before.

\section{An extension to ${\rm SU}(N)$ \label{sec4}}

The extension to the non-Abelian group ${\rm SU}(N)$ is straightforward without much modification, and our notations for the discussion of ${\rm U}(1)$ case in the previous section are already chosen for them to be applicable to ${\rm SU}(N)$, e.g., our use of $\phi_{\rm a}^*$, as long as we understand these fields as matrix-valued fields taking values in the ${\rm SU}(N)$ algebra. As $\phi_{\rm a}^\dagger$ and $\phi_{\rm a}^*$ are different for the non-Abelian fields, we will also use in this section the hermitian conjugates $\phi_{\rm a}^\dagger$ and $m_{\rm a}^\dagger$, whose KMS transformations are
\be
\phi_{\rm a}^\dagger(t)\to -\phi_{\rm a}^{\rm T}(-t)+\beta U_{\rm r}^{\rm T}(\partial_t U_{\rm r}^*)(-t)\,,\quad
m_{\rm a}^\dagger(t)\to -m_{\rm a}^{\rm T}(-t)+\beta M_{\rm r}^{\rm T}\left[\partial_t(M_{\rm r}^{\rm T})^{-1}\right](-t)\,.
\ee

The ${\rm SU}(N)$ symmetric action up to quadratic in $B$-fields is the same as in the ${\rm U}(1)$ case (\ref{eq:L0-U(1)}), with additional trace applied to all terms,
\be
\mathcal{L}_{\mathrm{SU}(N)}^{(2)} =  2\,{\rm tr}\left[\chi B_{{\rm a}t}B_{{\rm r}t}-f_\pi^2 B_{{\rm a}i}B_{{\rm r}i}-\chi^2\kappa B_{{\rm a}t}\left(\partial_t B_{{\rm r}t}-{\rm i}\beta^{-1}B_{{\rm a}t}\right)-\lambda B_{{\rm a}i}\left(\partial_t B_{{\rm r}i}-{\rm i}\beta^{-1}B_{{\rm a}i}\right)\right]\,,
\ee
where we use the fundamental representation with ${\rm tr}(t_at_b)=\delta_{ab}/2$. Each term is invariant under the full shift symmetry, except the second term, which is allowed only when there is no shift symmetry, i.e., the ${\rm SU}(N)$ symmetry is spontaneously broken. 

More general structures are possible for all terms: different $t_a$ may have different values of $\chi$, $f_\pi^2$, and the kinetic coefficients, $\kappa$ and $\lambda$. 
For this general case, $\chi$ and $f_\pi^2$, etc., become a matrix, depending on the detail of the SSB pattern. Moreover, if the symmetry is partially broken spontaneously, i.e., ${\rm SU}(N)\to H$, then the restricted shift symmetry in $H$ may require other structures, e.g., $2{\rm tr}(f_\pi^2 B_{{\rm a}i}PB_{{\rm r}i})$, where $P$ is the projection operator to the subspace ${\rm SU}(N)/H$ in the group algebra, so that the propagating NG bosons exist only for ${\rm SU}(N)/H$. We will not discuss these general possibilities and consider only the simplest case with $H\ni \{\emptyset\}$ in this section. We consider nontrivial $H$ for a specific case of chiral symmetry breaking in the next section.

We define the matrix valued chemical potential as $\mu=U_{\rm r}^\dagger B_{{\rm r}t}U_{\rm r}={\rm i}(D_t U_{\rm r}^\dagger)U_{\rm r}$, which can also be written in terms of generators as $\mu=\mu_a t_a$, and $\mu_a=2\,{\rm tr}(t_a\mu)$. Note that we distinguish the subscripts  ``a" and ``$a$" for the r- and a-type fields and the ${\rm SU}(N)$ index introduced here, respectively. Recall that (\ref{eq:Br,Ba}) and (\ref{eq:covariant-phi_a}), we take the variations with respect to $A_{{\rm a}\mu}$ and find the matrix-valued current,
\be
J^t_{\rm r}=\chi \mu-\chi^2\kappa D_t\mu+\xi_\kappa,\quad J^i_{\rm r}=-f_\pi^2 U_{\rm r}^\dagger B_{{\rm r}i}U_{\rm r}-\lambda(D_i\mu-E_i)+\xi^i\,,
\ee
where $D_\mu \mu=\partial_\mu\mu-{\rm i}[A_{{\rm r}\mu},\mu]$ and $E_i=\partial_t A_{ri}-\partial_i A_{rt}-{\rm i}[A_{rt},A_{ri}]$, and we used $[D_t,D_i]U_{\rm r}^\dagger=-iE_iU_{\rm r}^\dagger$. The noise correlations are
\be
\langle \xi_\kappa^a(x)\xi_\kappa^b(x')\rangle=2\chi^2 \kappa T\delta^{ab}\delta^{(4)}(x-x')\,,\quad
\langle \xi^{ai}(x)\xi^{bj}(x')\rangle=2\lambda \beta^{-1} \delta^{ab}\delta^{ij}\delta^{(4)}(x-x')\,,
\ee
where $\xi_\kappa=\xi_\kappa^a t_a$ and $\xi^i=\xi^{ia}t_a$. The EOM for the classical field takes the form of the conservation law,
\be
D_\mu j^\mu_r\equiv\partial_\mu j^\mu_r-{\rm i}[A_{{\rm r}\mu},j^\mu_r]=0\,.
\ee
Note that the gauge-dependent term in the covariant derivative arises from the gauge-field dependence in $B_{\rm a \mu}$ through (\ref{eq:covariant-phi_a}).

In the phase where the symmetry is not spontaneously broken, we have $f_\pi^2=0$, and each $\mu_a$ is the independent dynamical degree of freedom, and we have a non-Abelian version of charge diffusion dynamics, where external non-Abelian gauge fields couple different $\mu_a$'s. Readers who are interested in the case shall refer to Ref.~\cite{Glorioso:2020loc}. On the other hand, in the phase where the symmetry is spontaneously broken,
we write $U_{\rm r}={\rm e}^{{\rm i}\phi_{\rm r}}$ in terms of the NG modes $\phi_{\rm r}=\phi_{\rm r}^a t_a$, which become propagating modes. The linearized EOM for $\phi_{\rm r}^a$ without external gauge fields and the dispersion relation is identical to the ${\rm U}(1)$ case for each generator $t_a$ independently, but a non-Abelian external gauge field would couple them together.

We introduce an explicit symmetry-breaking parameter $M$, which we assume is a matrix transforming in the same representation of the $U$ field as given by (\ref{eq:M-sperious-gauge-trans}). The KMS transformation is what is given in the previous section, Eqs.~(\ref{eq:KMS-Mrma}) and (\ref{eq:KMS-mastar}), whereas in the present case, the matrix structure of $M$ is understood. In first order in $M_{\rm r}$, it can be shown by a similar argument in the ${\rm U}(1)$ case that there is no action that is consistent with the shift symmetry, and the effect of $M_{\rm r}$ in the phase where the symmetry is not spontaneously broken appears in the second order in $M_{\rm r}$.

In the SSB phase, the unique KMS invariant action at leading derivatives is as before (\ref{eq:L_mass-linear}),
\be 
{\cal L}_M=-{{\rm i}\sigma} \,{\rm tr} \mathcal{M}_{\rm a} +{\rm c.c.}\simeq -2\sigma M\,{\rm tr}(\phi_{\rm a} \phi_{\rm r})\,,
\ee
where $\mathcal{M}_{\rm a}$ is given by (\ref{eq:calMa}) and we assume a physical value $M_{\rm r}=M{\bf 1}$ and linearlized the fields in the last expression.
The EOM is modified as
\be
D_\mu j^\mu_r=-{{\rm i}\sigma\over 2}(U_{\rm r}^\dagger M_{\rm r}-M_{\rm r}^\dagger U_{\rm r})\simeq  -\sigma M \phi_{\rm r}
\equiv -\chi m_\pi^2 \phi_{\rm r}\,,
\ee
with the NG boson mass $m_\pi^2=\sigma M/\chi$. The linearized dispersion relation for each $\phi^a_{\rm r}$ is again identical to the ${\rm U}(1)$ case (\ref{eq:dispersion-U(1)}).

We next discuss possible terms in the second order of $M_{\rm r}$. Following the same steps in Sec.~\ref{sec:mass-second-order} for the U(1) case, we consider a KMS invariant term ${\rm tr}\left[ F(KF) \right]$, with the most general form of $F$,
\be
\label{eq:F-mass}
F=c_1\mathcal{M}_{\rm a}+c_2\mathcal{M}_{\rm a}^{\rm T}+c_3\mathcal{M}_{\rm a}^*+c_4\mathcal{M}_{\rm a}^\dagger\,,
\ee
where $\mathcal{M}_{\rm a}$ as well as $\mathcal{M}_{\rm a}^{\rm T}$, $\mathcal{M}_{\rm a}^*$, $\mathcal{M}_{\rm a}^\dagger$, are all gauge-invariant matrices.\footnote{One can also consider a KMS invariant term of the form, ${\rm tr}(F){\rm tr}(KF)$. However, due to ${\rm tr}(t_a)=0$ for ${\rm SU}(N)$, this structure does not produce the interesting terms responsible for a linearized charge relaxation.}
We have the KMS transformation for each operator:
\begin{subequations}
\label{eq:KMS-Ma}
\begin{align}
K\mathcal{M}_{\rm a}&=-\mathcal{M}_{\rm a}^*-\beta\partial_t(M_{\rm r}^*U_{\rm r}^{\rm T})\,,\quad
K\mathcal{M}_{\rm a}^{\rm T}= -\mathcal{M}_{\rm a}^\dagger-\beta\partial_t(U_{\rm r}M_{\rm r}^\dagger)\,,\\
K\mathcal{M}_{\rm a}^*&= -\mathcal{M}_{\rm a} +\beta\partial_t(M_{\rm r}U_{\rm r}^\dagger)\,,\quad
K\mathcal{M}_{\rm a}^\dagger= -\mathcal{M}_{\rm a}^{\rm T}+\beta\partial_t(U_{\rm r}^*M_{\rm r}^{\rm T})\,.
\end{align}
\end{subequations}
It can be shown that ${\rm tr}\left[ F(KF) \right]$ is a linear combination of the following six independent KMS invariant terms consistent with unitarity and stability,
\begin{subequations}
\begin{align}
\label{eq:Ls}
L_1&=-{{\rm i}\over 2}{\rm tr} \left[ \mathcal{M}_{\rm a} (K\mathcal{M}_{\rm a}^{\rm T})+\mathcal{M}_{\rm a}^\dagger (K\mathcal{M}_{\rm a}^*) \right]\,,\\
L_2&=-{{\rm i}\over 2}{\rm tr}\left[ \mathcal{M}_{\rm a}(K\mathcal{M}_{\rm a}^*)+\mathcal{M}_{\rm a}^\dagger (K\mathcal{M}_{\rm a}^{\rm T}) \right]\,,\\
L_3&=-\beta^{-1}{\rm tr}\left[ \mathcal{M}_{\rm a} (K\mathcal{M}_{\rm a}^{\rm T})-\mathcal{M}_{\rm a}^\dagger K\mathcal{M}_{\rm a}^*\right]\,,\\
L_4&=-{{\rm i}\over 2}{\rm tr}\left[ \mathcal{M}_{\rm a} (K\mathcal{M}_{\rm a})+\mathcal{M}_{\rm a}^\dagger (K\mathcal{M}_{\rm a}^\dagger)\right]\,,\\
L_5&=-{{\rm i}\over 2}{\rm tr}\left[ \mathcal{M}_{\rm a} (K\mathcal{M}_{\rm a}^\dagger)+\mathcal{M}_{\rm a}^\dagger (K\mathcal{M}_{\rm a})\right]\,,\\
L_6&=-\beta^{-1}{\rm tr}\left[ \mathcal{M}_{\rm a}(K\mathcal{M}_{\rm a})-\mathcal{M}_{\rm a}^\dagger (K\mathcal{M}_{\rm a}^\dagger) \right],\label{KMSinvariants}
\end{align}
\end{subequations}
which means that a general KMS invariant term satisfying the unitarity and stability conditions is a linear combination of $L_n$'s with positive real coefficients.
The $L_3$ and $L_6$ are purely real and non-dissipative,
\begin{subequations}
\begin{align}
L_3&={\rm tr}\left[\mathcal{M}_{\rm a}\partial_t(U_{\rm r}M_{\rm r}^\dagger)+\mathcal{M}_{\rm a}^\dagger\partial_t(M_{\rm r}U_{\rm r}^\dagger)\right]\,,\\
L_6&={\rm tr}\left[\mathcal{M}_{\rm a}\partial_t(M_{\rm r}^*U_{\rm r}^{\rm T})+\mathcal{M}_{\rm a}^\dagger\partial_t(U_{\rm r}^*M_{\rm r}^{\rm T})\right]\,,
\end{align}
\end{subequations}
which are higher orders in $M$ and $\partial_t$ relative to $\mathcal L_M$, so that we will not discuss about them.
Among the remaining dissipative terms, only $L_1$ is invariant under the full shift symmetry, which transforms $U_{\rm r}\to \Lambda(\bm x) U_{\rm r}$ and $\mathcal{M}_{\rm a}\to \mathcal{M}_{\rm a}\Lambda^\dagger(\bm x)$,
\be
L_1={\rm tr}\left\{{\rm i}\mathcal{M}_{\rm a}\mathcal{M}_{\rm a}^\dagger+{{\rm i}\beta\over 2}\left[\mathcal{M}_{\rm a}\partial_t(U_{\rm r}M_{\rm r}^\dagger)-\mathcal{M}_{\rm a}^\dagger\partial_t(M_{\rm r}U_{\rm r}^\dagger)\right]\right\}\,.
\ee
The next interesting term is $L_2$,
\be
L_2={\rm tr}\left\{ {{\rm i}\over 2}\left[\mathcal{M}_{\rm a}^2+(\mathcal{M}_{\rm a}^\dagger)^2\right]-{{\rm i}\beta\over 2}\left[\mathcal{M}_{\rm a}\partial_t(M_{\rm r}U_{\rm r}^\dagger)-\mathcal{M}_{\rm a}^\dagger\partial_t(U_{\rm r}M_{\rm r}^\dagger)\right]\right\}\,,
\ee
which is not invariant under the shift symmetry, but its effect is independent of the generator $t_a$, i.e., all $\phi^a_{\rm r}$ receive the same effect from this term.
On the other hand, $L_4$ and $L_5$, which contain terms that are proportional to ${\rm tr}[\mathcal{M}_{\rm a}(\mathcal{M}_{\rm a})^*]$ and ${\rm tr}[\mathcal{M}_{\rm a}(\mathcal{M}_{\rm a})^{\rm T}]$, respectively, give in general different contributions for different group directions since $t_a^*=t_a^{\rm T}$ is not universally related to $t_a$, unless the representation is real, e.g., ${\rm SU}(2)$ where $t_a=\sigma_a/2$ and $t_a^*=-\sigma_2 t_a\sigma_2$.
These terms may be possible if the pattern of the SSB is such as to allow a non-universality among different group directions, but we will consider only $L_1$ and $L_2$ here for simplicity.

In the phase where the symmetry is not spontaneously broken, we have only $L_1$ in the action, and using the condition $D_t M_{\rm r}=0$ for physical value, we find
\be
2\gamma_1\beta^{-1}L_1=-{\gamma_1}{\rm tr}\left(\phi_{\rm a} \{M_{\rm r}^\dagger M_{\rm r},\mu\}\right)+2{\rm i}\gamma_1\beta^{-1}{\rm tr}\left(\phi_{\rm a}^2M_{\rm r}^\dagger M_{\rm r}\right)\,,
\ee
where $\{A,B\}=AB+BA$. The EOM from this is
\be
D_\mu j^\mu_r=-{\gamma_1\over 2}\{M_{\rm r}^\dagger M_{\rm r},\mu\}+\xi\,,
\ee
which describes stochastic relaxation dynamics for the dynamical degree of freedom $\mu$, where the noise $\xi=\xi^a t_a$ has the correlation
\be
\langle\xi^a(x)\xi^b(x')\rangle=2\gamma_1 T\,{\rm tr}\left( \{t_a,t_b\}M_{\rm r}^\dagger M_{\rm r} \right)\delta^{(4)}(x-x')\,.
\ee

In the SSB phase, the $L_2$ term is also present,
\begin{align}
2\gamma_2\beta^{-1}L_2 &=-\gamma_2{\rm tr}\left[\phi_{\rm a}(U_{\rm r}^\dagger M_{\rm r} \mu U_{\rm r}^\dagger M_{\rm r}+M_{\rm r}^\dagger U_{\rm r}\mu M_{\rm r}^\dagger U_{\rm r})\right]\nonumber\\
&\quad+{\rm i}\gamma_2 \beta^{-1} {\rm tr}\left(\phi_{\rm a}U_{\rm r}^\dagger M_{\rm r}\phi_{\rm a}U_{\rm r}^\dagger M_{\rm r}+\phi_{\rm a} M_{\rm r}^\dagger U_{\rm r}\phi_{\rm a}M_{\rm r}^\dagger U_{\rm r}\right)\,.
\end{align}
The stability condition gives a constraint $\gamma_1>|\gamma_2|$, and the EOM is
\be
D_\mu j^\mu_r=-{{\rm i}\sigma\over 2}(M_{\rm r}U_{\rm r}^\dagger-U_{\rm r}M_{\rm r}^\dagger)-{\gamma_1\over 2}\{M_{\rm r}^\dagger M_{\rm r},\mu\}-{\gamma_2\over 2}\left(U_{\rm r}^\dagger M_{\rm r} \mu U_{\rm r}^\dagger M_{\rm r}+M_{\rm r}^\dagger U_{\rm r}\mu M_{\rm r}^\dagger U_{\rm r}\right)+\xi\,,
\ee
with the noise $\xi=\xi^a t_a$, where
\be
 \begin{split}
 \langle\xi^a(x)\xi^b(x')\rangle
 =2T
 \big[&\gamma_1{\rm tr}(\{t_a,t_b\}M_{\rm r}^\dagger M_{\rm r})
 \\
 &+{\gamma_2}{\rm tr}(t_a U_{\rm r}^\dagger M_{\rm r}t_bU_{\rm r}^\dagger M_{\rm r}+t_a M_{\rm r}^\dagger U_{\rm r}t_bM_{\rm r}^\dagger U_{\rm r})\big]\delta^{(4)}(x-x').
 \end{split}
\ee
If we assume $M_{\rm r}=M {\bf 1}$, the linearized dispersion relation for $\phi_{\rm r}^a$ is independent of $t_a$ and has precisely the same form as in the ${\rm U}(1)$ case, (\ref{eq:dispersion-U(1)-SSB}).

\section{The case of chiral symmetry ${\rm SU}(N)_{\rm L}\times {\rm SU}(N)_{\rm R}$\label{chiral} }

We now discuss our main application of the developed formalism: the chiral symmetry of QCD, ${\rm SU}(N)_{\rm L}\times {\rm SU}(N)_{\rm R}$. We introduce the two group fields, $U_{\rm L}$ and $U_{\rm R}$, which transform under the chiral symmetry as 
\be
U_{\rm L}\to U_{\rm L} V_{\rm L}^\dagger\,,\quad U_{\rm R}\to U_{\rm R} V_{\rm R}^\dagger\,,
\ee
where $V_{\rm L/R}\in {\rm SU}(N)_{\rm L/R}$. 
In the classical limit, we have the r- and a-type fields, $(U_{\rm Lr}\,,\ U_{\rm Rr})$ and $(\phi_{\rm La}\,,\ \phi_{\rm Ra})$, respectively for each ${\rm SU}(N)_{\rm L}$ and ${\rm SU}(N)_{\rm R}$. These are simply two copies of the ${\rm SU}(N)$ case in the previous section. We also have the external gauge fields $A_{{\rm L}\mu}$ and $A_{{\rm R}\mu}$, and the gauge invariant $B$-fields as before,
\be
B_{{\rm Lr}\mu}={\rm i} U_{\rm Lr}(D_\mu U_{\rm Lr}^\dagger)=-{\rm i}(D_\mu U_{\rm Lr})U_{\rm Lr}^\dagger\,, \quad B_{{\rm La}\mu}=U_{\rm Lr}(D_\mu \phi_{\rm La})U_{\rm Lr}^\dagger\,,
\ee
and similarly for $B_{{\rm Rr}\mu}$ and $B_{{\rm Ra}\mu}$. Recall that the gauge transformations for the group fields are
 \be
 U_{\rm Lr}\to U_{\rm Lr}V_{\rm Lr}^\dagger\,,\quad \phi_{\rm La}\to V_{\rm Lr}(\phi_{\rm La}-v_{\rm La})V_{\rm Lr}^\dagger\,,
 \ee
 and similarly for $U_{\rm Rr}$ and $\phi_{\rm Ra}$.

The explicit symmetry-breaking parameter we consider is the quark mass matrix $M$, which transforms under the chiral symmetry as 
\be
M\to V_{\rm L} M V_{\rm R}^\dagger\,,
\ee
and in the SK contour in the classical limit, we have the $r$ and $a$ fields, $(M_{\rm r},m_{\rm La},m_{\rm Ra})$, defined by
\be
M_1=\left(1-{{\rm i}\hbar\over 2}m_{\rm La}\right)M_{\rm r} \left(1+{{\rm i}\hbar\over 2}m_{\rm Ra}\right)+\cdots\,,\quad
M_2=\left(1+{{\rm i}\hbar\over 2}m_{\rm La}\right)M_{\rm r} \left(1-{{\rm i}\hbar\over 2}m_{\rm Ra}\right)+\cdots\,.
\ee
The $(m_{\rm La},m_{\rm Ra})$ are not independent, and only the combination $(m_{\rm La}M_{\rm r}-M_{\rm r} m_{\rm Ra})$ is meaningful, and the action we construct later will contain only this combination. This way of splitting will be useful to see the gauge invariance structure of the action terms later, as the gauge transformation can be chosen to be
\be
M_{\rm r}\to V_{\rm Lr}M_{\rm r}V_{\rm Rr}^\dagger\,, \quad m_{\rm La}\to V_{\rm Lr}(m_{\rm La}-v_{\rm La})V_{\rm Lr}^\dagger\,, \quad m_{\rm Ra}\to V_{\rm Rr}(m_{\rm Ra}-v_{\rm Ra})V_{\rm Rr}^\dagger\,.
\ee
The KMS transformation for the $M$-fields is $M_{\rm r}(t)\to M_{\rm r}^*(-t)$ and
\be
m_{\rm La}(t)\to -m_{\rm La}^*(-t)-{\beta\over 2}(\partial_t M_{\rm r}^*)(M_{\rm r}^*)^{-1}(-t)\,, \quad m_{\rm Ra}(t)\to -m_{\rm Ra}^*(-t)+{\beta\over 2}(M_{\rm r}^*)^{-1}(\partial_t M_{\rm r}^*)(-t)\,,
\ee
while the KMS transformation for $U_{\rm L/R}$-fields is the same as in the previous section,
\be
U_{\rm Lr}(t)\to U_{\rm Lr}^*(-t)\,,\quad \phi_{\rm La}(t)\to -\phi_{\rm La}^*(-t)+\beta U_{\rm Lr}^{\rm T}(\partial_t U_{\rm Lr}^*)(-t)\,,
 \ee
 and similarly for the right-handed fields.

We first discuss the symmetric fluid at the zero'th order in $M_{\rm r}$. If the chiral symmetry is not spontaneously broken, we impose the full ${\rm SU}(N)_{\rm L}\times {\rm SU}(N)_{\rm R}$ chiral shift symmetry, 
\be
U_{\rm Lr}\to \Lambda_{\rm L} U_{\rm Lr}\,,\quad U_{\rm Rr}\to \Lambda_{\rm R}U_{\rm Rr}\,,
\ee
and the action at quadratic order in derivatives should be a simple sum of two independent ${\rm SU}(N)$ actions in the previous section, and the dynamics of left and right chiral symmetries decouple.
Since the QCD interactions are identical for the left- and right-handed quarks, we additionally impose a $\mathbb{Z}_2$ symmetry on the action between $U_{\rm L}\leftrightarrow U_{\rm R}$. Then, the parameters in the action for the left- and right-handed chiral symmetries should be identical to each other.

The more interesting case is the phase with the spontaneous breaking to the vector part of the chiral symmetry. To figure out the corresponding partial shift symmetry we need to impose on the action, it is helpful to think of the condensate responsible for the symmetry breaking as being proportional to the group field $\Sigma \equiv U_{\rm L}^\dagger U_{\rm R}$, which transforms under the chiral symmetry as the usual group field in the chiral effective theory
\be
\Sigma\to V_{\rm L} \Sigma V_{\rm R}^\dagger\,.
\ee
We want our effective theory to contain similar terms that appear in the chiral effective theory constructed in terms of $\Sigma$, which means that $\Sigma$ should be invariant under the partial shift symmetry we impose on the fields. This fixes the partial shift symmetry to be the vector-like one with $\Lambda_{\rm L}=\Lambda_{\rm R}=\Lambda(\bm x)$, i.e., $U_{\rm L/R}\to \Lambda(\bm x) U_{\rm L/R}$ with the same $\Lambda(\bm x)$, under which $\Sigma$ is indeed invariant. The gauge invariant $B$-fields transform under the shift symmetry as
\begin{align}
(B_{{\rm Lr}t}\,,\ B_{{\rm Rr}t}\,,\ B_{{\rm La}\mu}\,,\ B_{{\rm Ra}\mu})&\to \Lambda(\bm x) (B_{{\rm Lr}t}\,,\ B_{{\rm Rr}t}\,,\ B_{{\rm La}\mu}\,,\ B_{{\rm Ra}\mu})\Lambda^\dagger(\bm x)\,, \\
(B_{{\rm Lr}i}\,,\ B_{{\rm Rr}i})& \to \Lambda(\bm x)(B_{{\rm Lr}i}\,,\ B_{{\rm Rr}i})\Lambda^\dagger(\bm x)+{\rm i}\Lambda(\bm x)\partial_i\Lambda^\dagger(\bm x)\,.
\end{align}
Note that the axial $B$-field$, B_{{\rm Ar}i} \equiv B_{{\rm Lr}i}-B_{{\rm Rr}i}={\rm i} U_{\rm Lr} D_i \Sigma U_{\rm Rr}^\dagger$, transforms covariantly under the shift symmetry, and the most general action respecting all symmetries is
\begin{align}
\mathcal{L}_0&={1\over 2}\,{\rm tr} \left[ \sum_{\alpha={\rm V,A}}\chi_{\alpha} B_{\alpha {\rm a}t}B_{\alpha {\rm r}t}-f_\pi^2 B_{{\rm Aa}i}B_{{\rm Ar}i} \right. \notag\\ 
& \left. \quad -\sum_{\alpha={\rm V},{\rm A}}\chi^2_{\alpha}\kappa_{\alpha} B_{\alpha {\rm a}t}\left(\partial_t B_{\alpha {\rm r}t}-{\rm i}\beta^{-1}B_{\alpha {\rm a}t}\right)-\sum_{\alpha={\rm V},{\rm A}}\lambda_{\alpha} B_{\alpha {\rm a}i}\left(\partial_t B_{\alpha {\rm r}i}-{\rm i}\beta^{-1}B_{\alpha {\rm a}i}\right) \right] \,,
\end{align}
where we also defined $B_{{\rm Aa}\mu}\equiv B_{{\rm La}\mu}-B_{{\rm Ra}\mu}$, and the vector-like $B$-fields, $B_{{\rm Vr}\mu}\equiv B_{{\rm Lr}\mu} + B_{{\rm Rr}\mu}$, $B_{{\rm Va}\mu}\equiv B_{{\rm La}\mu}+B_{{\rm Ra}\mu}$, and $\chi_{\rm V/A}$ are vector/axial susceptibilities, respectively.
The vector-like symmetry sector remains as a simple charge diffusion system, while the axial sector, where $f_\pi^2$-term is non-zero, is described by the propagating NG pions, $U_{\rm Lr}^\dagger=U_{\rm Rr}={\rm e}^{{\rm i}\phi_{\rm r}}$, with a velocity $v_{\rm s}^2$. The linearized dispersion relation for $\phi_{\rm r}=\phi^a_{\rm r} t_a$ is identical to the ${\rm U}(1)$ and ${\rm SU}(N)$ case in the previous sections. The axial part of the action can also be related to the more conventional chiral field $\Sigma=U_{\rm L}^\dagger U_{\rm R}$. Expanding $\Sigma_{1,2}=\Sigma_{\rm r}\pm{{\rm i}\hbar/2}(\Sigma_{\rm r} \phi_{\rm Ra}-\phi_{\rm La}\Sigma_{\rm r})\equiv \Sigma_{\rm r}\pm\hbar/2\Sigma_{\rm a}$ with $\Sigma_{\rm r}=U_{\rm Lr}^\dagger U_{\rm Rr}$, it can be shown that the classical limit of the conventional chiral effective theory reduces to the leading derivative term of the axial part in the above action ($v_{\rm s} = 1$),
\be
\lim_{\hbar\to 0}{1\over \hbar}\left[{\rm tr}\left(D_\mu \Sigma_1^\dagger D_\mu\Sigma_1-D_\mu \Sigma_2^\dagger D_\mu\Sigma_2\right)\right]={\rm tr}\left(D_\mu \Sigma_{\rm r}^\dagger D_\mu\Sigma_{\rm a}+D_\mu\Sigma_{\rm a}^\dagger D_\mu \Sigma_{\rm r}\right)={\rm tr}\left(B_{{\rm Aa}\mu}B_{{\rm Ar}\mu}\right)\,.
\ee

We define the chiral chemical potentials as in the ${\rm SU}(N)$ case,
\be
\mu_{\rm L}=U_{\rm Lr}^\dagger B_{{\rm Lr}t}U_{\rm Lr}={\rm i}D_t U_{\rm Lr}^\dagger U_{\rm Lr} \,,
\ee
similarly for $\mu_{\rm R}$. Note that $\Sigma_{\rm r}^\dagger D_t\Sigma_{\rm r} =-{\rm i}(\Sigma_{\rm r}^\dagger \mu_{\rm L}\Sigma_{\rm r} -\mu_{\rm R})$ and $D_t\Sigma_{\rm r} \Sigma_{\rm r}^\dagger=-{\rm i}(\mu_{\rm L}-\Sigma_{\rm r}\mu_{\rm R}\Sigma_{\rm r}^\dagger)$.
The chiral currents obtained from variations with respect to external gauge fields are
\begin{subequations}
\begin{align}
j_{\rm Lr}^t&= {1\over 4}\Big[\chi_{\rm V} (\mu_{\rm L}+\Sigma_{\rm r}\mu_{\rm R}\Sigma_{\rm r}^\dagger)+\chi_{\rm A}(\mu_{\rm L}-\Sigma_{\rm r}\mu_{\rm R}\Sigma_{\rm r}^\dagger)\nonumber\\
&\quad -\chi_{\rm V}^2\kappa_{\rm V}(D_t\mu_{\rm L}+\Sigma_{\rm r} D_t\mu_{\rm R} \Sigma_{\rm r}^\dagger)-\chi_{\rm A}^2\kappa_{\rm A}(D_t\mu_{\rm L}-\Sigma_{\rm r} D_t\mu_{\rm R} \Sigma_{\rm r}^\dagger)\Big]+{1\over 2}U_{\rm Lr}^\dagger(\xi_{\kappa {\rm V}}+\xi_{\kappa {\rm A}})U_{\rm Lr}\,, \\
%%%
j_{\rm Rr}^t&={1\over 4}\Big[\chi_{\rm V} (\mu_{\rm R}+\Sigma_{\rm r}^\dagger\mu_{\rm L}\Sigma_{\rm r})+\chi_{\rm A}(\mu_{\rm R}-\Sigma_{\rm r}^\dagger\mu_{\rm L}\Sigma_{\rm r})\nonumber\\
&\quad -\chi_{\rm V}^2\kappa_{\rm V}(D_t\mu_{\rm R}+\Sigma_{\rm r}^\dagger D_t\mu_{\rm L} \Sigma_{\rm r})-\chi_{\rm A}^2\kappa_{\rm A}(D_t\mu_{\rm R}-\Sigma_{\rm r}^\dagger D_t\mu_{\rm L} \Sigma_{\rm r})\Big]+{1\over 2}U_{\rm Rr}^\dagger(\xi_{\kappa {\rm V}}-\xi_{\kappa {\rm A}})U_{\rm Rr}\,,\\
%%%
J_{\rm Lr}^i&= {1\over 4}\Big[{\rm i}f_\pi^2 \Sigma_{\rm r} D_i \Sigma_{\rm r}^\dagger -\lambda_{\rm V}\left[D_i\mu_{\rm L}-E_{{\rm L}i}+\Sigma_{\rm r}(D_i\mu_{\rm R}-E_{{\rm R}i})\Sigma_{\rm r}^\dagger\right]\nonumber\\
&\quad -\lambda_{\rm A}\left[D_i\mu_{\rm L}-E_{{\rm L}i}-\Sigma_{\rm r}(D_i\mu_{\rm R}-E_{{\rm R}i})\Sigma_{\rm r}^\dagger\right]\Big]+{1\over 2}U_{\rm Lr}^\dagger (\xi_{\rm V}^i+\xi_{\rm A}^i)U_{\rm Lr}\,,\\
%%%
J_{\rm Rr}^i&= {1\over 4}\Big\{{\rm i}f_\pi^2 \Sigma_{\rm r}^\dagger  D_i \Sigma_{\rm r}-\lambda_{\rm V}\left[D_i\mu_{\rm R}-E_{{\rm R}i}+\Sigma_{\rm r}^\dagger(D_i\mu_{\rm L}-E_{{\rm L}i})\Sigma_{\rm r}\right]\nonumber\\
&\quad -\lambda_{\rm A}\left[D_i\mu_{\rm R}-E_{{\rm R}i}-\Sigma_{\rm r}^\dagger(D_i\mu_{\rm L}-E_{{\rm L}i})\Sigma_{\rm r}\right] \Big\}+{1\over 2}U_{\rm Rr}^\dagger (\xi_{\rm V}^i-\xi_{\rm A}^i)U_{\rm Rr}\,,
\end{align}
\end{subequations}
where the thermal noises, $\xi_{\kappa\alpha}=\xi^a_{\kappa\alpha}t_a$ and $\xi_\alpha^i=\xi_\alpha^{ia}t_a$ $(\alpha={\rm V,A})$, have the correlation functions,
\begin{subequations}
 \begin{align}
 \langle \xi^a_{\kappa\alpha} (x) \xi^b_{\kappa\alpha} (x') \rangle 
 &=2T\chi_\alpha^2\kappa_\alpha\delta^{ab} \delta^{(4)} (x-x')\,,
 \\
 \langle \xi^{ia}_\alpha (x) \xi^{jb}_\alpha (x') \rangle
 &=2T\lambda_\alpha\delta^{ab}\delta^{ij} \delta^{(4)} (x-x')\,,
 \end{align}
\end{subequations}
with $(\alpha={\rm V,A})$.
The EOM following from the variations of the effective action with respect to $\phi_{\rm La}$ and $\phi_{\rm Ra}$ are given by the conservation laws,
\be
D_\mu j^\mu_{{\rm L} {\rm r}}=0\,,
\ee
and likewise for the right current $j_{\rm R}$. 

We next discuss the linear term in $M_{\rm r}$, which exists only in the phase of spontaneously broken axial symmetry. It is the classical limit of the usual mass term in the chiral effective theory,
${\rm tr}(U_{\rm L} M U_{\rm R}^\dagger)+{\rm h.c.}={\rm tr}(M\Sigma^\dagger)+{\rm h.c.}$, which is put in the SK contour, 
\begin{align}
L_M&={1\over\hbar}\sigma\,{\rm tr}\left(M_1\Sigma_1^\dagger -M_2 \Sigma_2^\dagger\right)+{\rm h.c.} \notag\\
&=-{\rm i}\sigma\,{\rm tr}\left\{ \left[M_{\rm r}(\phi_{\rm Ra}-m_{\rm Ra})-(\phi_{\rm La}-m_{\rm La})M_{\rm r}\right] \Sigma_{\rm r}^\dagger\right\} +{\rm h.c.}.
\end{align}
Again, we emphasize that this term is excluded by the chiral shift symmetry when there is no spontaneous breaking of it. It is easily shown to be KMS invariant
up to the total derivative term
\be
-{\rm i}\sigma\beta\partial_{t}\left\{{\rm tr}[M_{\rm r}^\dagger(-t) \Sigma_{\rm r}(-t)]\right\}-{\rm h.c.}
\ee
The EOM is modified as
\be
D_\mu j^\mu_{\rm Lr}={{\rm i}\sigma\over 2}\left(M_{\rm r}\Sigma_{\rm r}^\dagger - \Sigma_{\rm r}M_{\rm r}^\dagger\right)\,,\quad
D_\mu j^\mu_{\rm Rr}={{\rm i}\sigma\over 2}\left(M_{\rm r}^\dagger\Sigma_{\rm r} - \Sigma_{\rm r}^\dagger M_{\rm r}\right)\,.
\ee
The linearized dispersion relation for the pions is again identical to the ${\rm SU}(N)$ case with the pion mass $m_\pi^2=4\sigma M/\chi_{\rm A}$, if $M_{\rm r}=M{\bf 1}$.

Finally, we construct the terms that are second order in $M_{\rm r}$. We find that the procedure and the outcome go parallel with those in the ${\rm SU}(N)$ case. The unique gauge invariant matrix $\mathcal{M}_{\rm a}$ that has the necessary properties we discussed in the previous section is
\be
\mathcal{M}_{\rm a}=U_{\rm Lr}\left[M_{\rm r}(\phi_{\rm Ra}-m_{\rm Ra})-(\phi_{\rm La}-m_{\rm La})M_{\rm r}\right] U_{\rm Rr}^\dagger\,,
\ee
and we consider a most general KMS-invariant structure in the action, ${\rm tr}(F (KF))$, where $F$ is a general gauge invariant matrix given by (\ref{eq:F-mass}). 
The K-transformation for $(\mathcal{M}_{\rm a},\mathcal{M}_{\rm a}^{\rm T},\mathcal{M}_{\rm a}^*,\mathcal{M}_{\rm a}^\dagger)$ has a very similar structure as in the ${\rm SU}(N)$ case (\ref{eq:KMS-Ma}), 
\begin{align}
K\mathcal{M}_{\rm a}&=-\mathcal{M}_{\rm a}^*-\beta\partial_t\left(U_{\rm Lr}^*M_{\rm r}^*U_{\rm Rr}^{\rm T}\right)\,,\quad
K\mathcal{M}_{\rm a}^{\rm T}= -\mathcal{M}_{\rm a}^\dagger-\beta\partial_t\left(U_{\rm Rr}M_{\rm r}^\dagger U_{\rm Lr}^\dagger\right),\\
K\mathcal{M}_{\rm a}^*&= -\mathcal{M}_{\rm a} +\beta\partial_t\left(U_{\rm Lr}M_{\rm r}U_{\rm Rr}^\dagger\right)\,,\quad
KF^\dagger= -\mathcal{M}_{\rm a}^{\rm T}+\beta\partial_t\left(U_{\rm Rr}^*M_{\rm r}^{\rm T}U_{\rm Lr}^{\rm T}\right).
\end{align}
and the most general KMS invariant term, which is consistent with the unitarity and stability conditions, is given by a linear combination of the six independent structures, $L_{\rm i}$, $i=1,\cdots,6$, whose expression in terms of $(\mathcal{M}_{\rm a},\mathcal{M}_{\rm a}^{\rm T},\mathcal{M}_{\rm a}^*,\mathcal{M}_{\rm a}^\dagger)$ takes the precisely same form as in the ${\rm SU}(N)$ case (\ref{eq:Ls}). Note that $\mathcal{M}_{\rm a}$ transforms under the full chiral shift symmetry as
\be
\mathcal{M}_{\rm a}\to \Lambda_{\rm L} \mathcal{M}_{\rm a} \Lambda_{\rm R}^\dagger,
\ee
and only $L_1$ is invariant under this, neglecting the non-dissipative $L_3$ and $L_6$,
\begin{align}
2\gamma_1\beta^{-1}L_1&=2\gamma_1\beta^{-1}{\rm tr}\left\{{\rm i}\mathcal{M}_{\rm a}\mathcal{M}_{\rm a}^\dagger+{{\rm i}\beta\over 2}\left[\mathcal{M}_{\rm a}\partial_t(U_{\rm Rr}M_{\rm r}^\dagger U_{\rm Lr}^\dagger) -\mathcal{M}_{\rm a}^\dagger\partial_t(U_{\rm Lr}M_{\rm r}U_{\rm Rr}^\dagger)\right]\right\} \nonumber\\
&=-\gamma_1{\rm tr}\left[\phi_{\rm La}\{M_{\rm r}M_{\rm r}^\dagger,\mu_{\rm L}\}+\phi_{\rm Ra}\{M_{\rm r}^\dagger M_{\rm r},\mu_{\rm R}\}-2(\phi_{\rm La}M_{\rm r}\mu_{\rm R}M_{\rm r}^\dagger +\phi_{\rm Ra}M_{\rm r}^\dagger \mu_{\rm L} M_{\rm r})\right]\nonumber\\
&\quad +2{\rm i}\gamma_1\beta^{-1}\,{\rm tr}\left(\phi_{\rm La}^2 M_{\rm r} M_{\rm r}^\dagger +\phi_{\rm Ra}^2M_{\rm r}^\dagger M_{\rm r} -2 \phi_{\rm La}M_{\rm r} \phi_{\rm Ra}M_{\rm r}^\dagger\right).
\end{align}
Therefore, in the phase where no chiral symmetry is spontaneously broken, only the $L_1$ is present, and it is the leading mechanism of axial charge relaxation.
The EOM takes the form
\begin{subequations}
\begin{align}
D_\mu j^\mu_{\rm Lr}&=-{\gamma_1\over 2}\left(\{M_{\rm r}M_{\rm r}^\dagger,\mu_{\rm L}\}-2M_{\rm r}\mu_{\rm R}M_{\rm r}^\dagger\right)+\xi_{\rm L}\,, \\
D_\mu j^\mu_{\rm Rr}&=-{\gamma_1\over 2}\left(\{M_{\rm r}^\dagger M_{\rm r},\mu_{\rm R}\}-2M_{\rm r}^\dagger\mu_{\rm L}M_{\rm r}\right) +\xi_{\rm R}\,,
\end{align}
\end{subequations}
with the thermal noise correlations
\begin{subequations}
\begin{align}
\langle \xi^a_{\rm L}(x)\xi^b_{\rm L}(x')\rangle&=2\gamma_1 T\, {\rm tr}\,(\{t_a,t_b\}M_{\rm r} M_{\rm r}^\dagger)\delta^{(4)}(x-x')\,, \\ \langle \xi^a_{\rm R}(x)\xi^b_{\rm R}(x')\rangle&=2\gamma_1 T\, {\rm tr}\, (\{t_a,t_b\}M_{\rm r}^\dagger M_{\rm r})\delta^{(4)}(x-x')\,,
\end{align}
\end{subequations}
and \be \langle \xi^a_{\rm L}(x)\xi^b_{\rm R}(x')\rangle=-4\gamma_1 T\,{\rm tr}\,(t_a M_{\rm r} t_b M_{\rm r}^\dagger)\delta^{(4)}(x-x')\,.
\ee
Recall that imposing the full chiral shift symmetry in this phase also dictates that the leading derivative terms in terms of $B$-fields decouple between the left and right-handed sectors, or equivalently we have $f_\pi^2=0$ and $(\chi_{\rm V}\,,\ \kappa_{\rm V}\,,\ \lambda_{\rm V})=(\chi_{\rm A}\,,\ \kappa_{\rm A}\,,\ \lambda_{\rm A})$. From these, the chiral currents are given only in terms of the chiral chemical potentials, without mixing between the left and right-handed sectors. Since the right-hand side of the EOM in the above also involves only the chiral chemical potentials without $\Sigma_{\rm r}$ field,
the hydrodynamic chiral dynamics in this phase is described completely in terms of the chiral chemical potentials $\mu_{\rm L/R}$.

On the other hand, in the phase where the axial symmetry is spontaneously broken and we impose only the vector-like shift symmetry, $\mathcal{M}_{\rm a}\to \Lambda \mathcal{M}_{\rm a} \Lambda^\dagger$, we find that $L_2$ is additionally allowed in the action, while $L_4$ and $L_5$ are still not consistent with the shift symmetry,
\begin{align}
2\gamma_2\beta^{-1}L_2&=2\gamma_2\beta^{-1}\left(-i\over 2\right){\rm tr}\left\{ -\mathcal{M}_{\rm a}^2-(\mathcal{M}_{\rm a}^\dagger)^2+\beta\left[\mathcal{M}_{\rm a}\partial_t(U_{\rm Lr}M_{\rm r}U_{\rm Rr}^\dagger)-\mathcal{M}_{\rm a}^\dagger\partial_t(U_{\rm Rr}M_{\rm r}^\dagger U_{\rm Lr}^\dagger)\right]\right\} \nonumber\\
&=-\gamma_2{\rm tr} \left[ (\phi_{\rm La}M_{\rm r}-M_{\rm r}\phi_{\rm Ra})\Sigma_{\rm r}^\dagger(\mu_{\rm L}M_{\rm r}-M_{\rm r}\mu_{\rm R})\Sigma_{\rm r}^\dagger +{\rm h.c.} \right]\nonumber\\
&\quad +{\rm i}\gamma_2\beta^{-1}\,{\rm tr}\left[ (M_{\rm r}^\dagger \phi_{\rm La}-\phi_{\rm Ra}M_{\rm r}^\dagger)\Sigma_{\rm r}(M_{\rm r}^\dagger \phi_{\rm La}-\phi_{\rm Ra}M_{\rm r}^\dagger)\Sigma_{\rm r}+{\rm h.c.}\right].
\end{align} 
The full EOM in this phase is
\begin{subequations}
\begin{align}
D_\mu j^\mu_{\rm Lr}&={{\rm i}\sigma\over 2}\left(M_{\rm r}\Sigma_{\rm r}^\dagger - \Sigma_{\rm r}M_{\rm r}^\dagger\right)-{\gamma_1\over 2}\left(\{M_{\rm r}M_{\rm r}^\dagger,\mu_{\rm L}\}-2M_{\rm r}\mu_{\rm R}M_{\rm r}^\dagger\right)\nonumber\\
&\quad -{\gamma_2\over 2}\left[M_{\rm r}\Sigma_{\rm r}^\dagger(\mu_{\rm L}M_{\rm r}-M_{\rm r}\mu_{\rm R})\Sigma_{\rm r}^\dagger+\Sigma_{\rm r}(M_{\rm r}^\dagger \mu_{\rm L}-\mu_{\rm R} M_{\rm r}^\dagger)\Sigma_{\rm r} M_{\rm r}^\dagger\right]+\xi_{\rm L}\,, \\
%%%
D_\mu j^\mu_{\rm Rr}&={{\rm i}\sigma\over 2}\left(M_{\rm r}^\dagger\Sigma_{\rm r} - \Sigma_{\rm r}^\dagger M_{\rm r}\right)-{\gamma_1\over 2}\left(\{M_{\rm r}^\dagger M_{\rm r},\mu_{\rm R}\}-2M_{\rm r}^\dagger\mu_{\rm L}M_{\rm r}\right)\nonumber\\
&\quad -{\gamma_2\over 2}\left[M_{\rm r}^\dagger\Sigma_{\rm r}(\mu_{\rm R}M_{\rm r}^\dagger -M_{\rm r}^\dagger \mu_{\rm L})\Sigma_{\rm r}+\Sigma_{\rm r}^\dagger (M_{\rm r} \mu_{\rm R}-\mu_{\rm L} M_{\rm r})\Sigma_{\rm r}^\dagger M_{\rm r}\right] +\xi_{\rm R}\,,
\end{align}
\end{subequations}
and the noise correlation functions are
\begin{subequations}
\begin{align}
\langle \xi^a_{\rm L}(x)\xi^b_{\rm L}(x')\rangle &= 2T\big[\gamma_1 {\rm tr}(\{t_a,t_b\}M_{\rm r} M_{\rm r}^\dagger)
\nonumber \\
&\hspace{36pt}
+\gamma_2{\rm tr}(M_{\rm r}^\dagger t_a \Sigma_{\rm r}M_{\rm r}^\dagger t_b\Sigma_{\rm r}+\Sigma_{\rm r}^\dagger t_a M_{\rm r}\Sigma_{\rm r}^\dagger t_b M_{\rm r})\big]\delta^{(4)}(x-x')\,,\\
%%%
\langle \xi^a_{\rm R} (x)\xi^b_{\rm R}(x')\rangle &= 2T\big[\gamma_1 {\rm tr}(\{t_a,t_b\}M_{\rm r}^\dagger M_{\rm r})
\nonumber \\
&\hspace{36pt}
+\gamma_2{\rm tr}(t_a M_{\rm r}^\dagger  \Sigma_{\rm r}t_b M_{\rm r}^\dagger \Sigma_{\rm r}+\Sigma_{\rm r}^\dagger M_{\rm r}t_a \Sigma_{\rm r}^\dagger M_{\rm r}t_b)\big]\delta^{(4)}(x-x')\,,\\
%%%
\langle \xi^a_{\rm L}(x)\xi^b_{\rm R}(x')\rangle &= -2T\big[2\gamma_1 {\rm tr}(t_aM_{\rm r} t_b M_{\rm r}^\dagger)
\nonumber \\
&\hspace{36pt}
+\gamma_2{\rm tr}(M_{\rm r}^\dagger t_a \Sigma_{\rm r} t_b M_{\rm r}^\dagger \Sigma_{\rm r}+\Sigma_{\rm r}^\dagger  M_{\rm r}t_b \Sigma_{\rm r}^\dagger t_a M_{\rm r})\big]\delta^{(4)}(x-x')\,.
\end{align}
\end{subequations}
Again, the stability condition dictates that $\gamma_1>|\gamma_2|$.

\section{Linearized dynamics and the Kubo formula \label{kubosection}} 

In this section, we study the linearized dynamics of the axial charge sector in the hydrodynamics of chiral symmetry (QCD), and derive the Kubo formula for the dissipative kinetic coefficients $\gamma_{1,2}$ that we introduced in the previous section. 
The linearized dynamics of axial and vector charges decouple, and the vector charge dynamics is a simple hydrodynamics diffusion of a conserved charge, so we will focus only on the axial charge dynamics which features a relaxation dynamics due to the explicit symmetry breaking by the quark mass matrix that we assume to be $M_{\rm r}=M{\bf 1}$ with the positive constant $M>0$.

For our purpose, we keep only the linearized fluctuations corresponding to the axial charge sector, and expand the fields in the absence of external gauge fields as 
\begin{align}
U_{\rm Lr}&={\rm e}^{-{\rm i}\phi_{\rm r}}\simeq 1-{\rm i}\phi_{\rm r}\,,\quad U_{\rm Rr}={\rm e}^{{\rm i}\phi_{\rm r}}\simeq 1+{\rm i}\phi_{\rm r}\,,\quad \Sigma_{\rm r}=U_{\rm Lr}^\dagger U_{\rm Rr}\simeq 1+2{\rm i}\phi_{\rm r}\,,\nonumber\\
\phi_{\rm La}&=-\phi_{\rm Ra}\equiv -\phi_{\rm a}\,,\quad B_{{\rm Ar}\mu}\simeq -2\partial_\mu\phi_{\rm r}\,,\quad B_{{\rm Aa}\mu}\simeq -2\partial_\mu\phi_{\rm a}\,,\quad \mu_{\rm A}\equiv (\mu_{\rm L}-\mu_{\rm R})/2\simeq -\partial_t\phi_{\rm r}\,.
\end{align}
It is sufficient to consider the quadratic action in $(\phi_{\rm r},\phi_{\rm a})$ for linearized dynamics.
The total effective Lagrangian is $\mathcal{L}=\mathcal{L}_0+\mathcal{L}_M+\mathcal{L}_\gamma$, where
\begin{subequations}
\begin{align}
\mathcal{L}_0&=2{\rm tr}\bigg[\chi_{\rm A} (\partial_t\phi_{\rm a}) (\partial_t\phi_{\rm r})-f_\pi^2 (\partial_i \phi_{\rm a})(\partial_{\rm i}\phi_{\rm r})
-\chi_{\rm A}^2\kappa_{\rm A} (\partial_t\phi_{\rm a})(\partial_t^2\phi_{\rm r}-{\rm i}\beta^{-1}\partial_t\phi_{\rm a})\nonumber\\
&\qquad \quad -\lambda_{\rm A} (\partial_i\phi_{\rm a})(\partial_i\partial_t\phi_{\rm r}-{\rm i}\beta^{-1}\partial_i\phi_{\rm a})\bigg]\,,\\
\mathcal{L}_M&=-8 \sigma M{\rm tr}(\phi_{\rm a} \phi_{\rm r})\,,
\\
\mathcal{L}_\gamma&= -8\gamma M^2 {\rm tr}\left(\phi_{\rm a} \partial_t\phi_{\rm r}\right)+8{\rm i}\gamma  M^2 \beta^{-1}{\rm tr}\left(\phi_{\rm a}^2\right)\,.
\end{align}
\end{subequations}
In the phase where there is no spontaneous breaking of axial symmetry, we have $f_\pi^2=\sigma=\gamma_2=0$, and the action can be written only in terms of $\phi_{\rm a}$ and $\mu_{\rm A}=-\partial_t\phi_{\rm r}$. In the phase where the axial symmetry is spontaneously broken, all the coefficients are non-vanishing in general, and the pseudo-NG bosons (pions), $\phi_{\rm r}$, become propagating dynamical modes. 
Note that $\gamma_1$ and $\gamma_2$ in this phase appear in the combination of $\gamma $ in the linearized dynamics, and they are distinguishable only in non-linear dynamics.

Introducing the random noise fields, $\xi_{\kappa {\rm A}}$, $\bm\xi_{\rm A}=\xi_{\rm A}^i$, and $\xi_{\gamma A}$, the $\mathcal{L}_0$ and $\mathcal{L}_\gamma$ can also be
written as
\begin{subequations}
\begin{align}
\mathcal{L}_0&=2{\rm tr}\bigg[\chi_{\rm A} (\partial_t\phi_{\rm a}) (\partial_t\phi_{\rm r})-f_\pi^2 (\partial_i \phi_{\rm a})(\partial_{\rm i}\phi_{\rm r})
-\chi_{\rm A}^2\kappa_{\rm A} (\partial_t\phi_{\rm a})(\partial_t^2\phi_{\rm r}) -\lambda_{\rm A} (\partial_i\phi_{\rm a})(\partial_i\partial_t\phi_{\rm r})\nonumber\\
& \qquad \quad -\xi_{\kappa {\rm A}}(\partial_t\phi_{\rm a})-\xi_{\rm A}^i (\partial_i\phi_{\rm a})+{{\rm i}\over 2\chi_{\rm A}^2\kappa_{\rm A} T}\xi_{\kappa {\rm A}}^2+{{\rm i}\over 2 \lambda_{\rm A} T}\xi_{\rm A}^i\xi_{\rm A}^i\bigg]\,,\\
\mathcal{L}_\gamma&= -8\gamma M^2 {\rm tr}\left(\phi_{\rm a} \partial_t\phi_{\rm r}\right)
-2{\rm tr}(\xi_{\gamma A}\phi_{\rm a})+{{\rm i}\over 8\gamma M^2 T}{\rm tr}(\xi_{\gamma A}^2)\,.
\end{align}
\end{subequations}
The path integral of the noise fields reproduces the original action, showing the equivalence.
The EOM for $\phi_{\rm r}$-field in this formalism takes the form of a stochastic hydrodynamics equation with random thermal noises. Another advantage is that the action becomes linear in the ``a"-type external gauge fields $A_{\rm a}$, and therefore there are no contact terms in the correlation functions of the ``r"-type charge currents $j_{\rm r}$ which is conjugate to $A_{\rm a}$; a feature that is shared by the microscopic QCD action.

The EOM can be written as (the violation of) the conservation Ward-Takahashi identity of the axial symmetry, 
\be
\partial_\mu j_{\rm A}^\mu=4\sigma M\phi_{\rm r} -4\gamma M^2 \mu_{\rm A}+\xi_{\gamma A}\equiv S_{\rm A}\,,
\ee
where the axial current is
\be
j_{\rm A}^t=\chi_{\rm A}\mu_{\rm A} -\chi_{\rm A}^2\kappa_{\rm A}\partial_t\mu_{\rm A}+\xi_{\kappa {\rm A}}\,,\quad j_{\rm A}^i=f_\pi^2\partial_{\rm i}\phi_{\rm r} -\lambda_{\rm A}\partial_i\mu_{\rm A}+\xi_{\rm A}^i\,,
\ee
and we defined the source operator, $S_{\rm A}$, for the axial charge violation. 
In underlying QCD, the source operator would correspond to $S_{\rm A}^a={\rm i}M\bar q\gamma^5 t_a q$, where $q$ is the quark Dirac field with the group index $a$. 
The noise correlations are given by 
\begin{subequations}
\begin{align}
\langle \xi^a_{\kappa {\rm A}}(x)\xi^b_{\kappa {\rm A}}(x')\rangle&=2T\chi_{\rm A}^2\kappa_{\rm A}\delta^{ab} \delta^{(4)}(x-x')\,,\\
\langle \xi_{\rm A}^{ia}(x)\xi_{\rm A}^{jb}(x')
\rangle&=2T\lambda_{\rm A}\delta^{ij}\delta^{ab}\delta^{(4)}(x-x')\,,\\\
 \langle \xi_{\gamma {\rm A}}^a(x)\xi_{\gamma {\rm A}}^b(x')\rangle&=8T\gamma M^2 \delta^{ab}\delta^{(4)}(x-x')\,.
\end{align}
\end{subequations}

In the phase where there is no SSB, the source-source fluctuation correlation function at zero spatial momentum is computed to be\footnote{We omit group index for simplicity. The left side is $\langle S_{\rm A}^a S_{\rm A}^b\rangle$ and the right side has $\delta^{ab}$ factor.}
\be
\langle S_{\rm A}(\omega)S_{\rm A}(-\omega)\rangle\equiv G^{S_{\rm A}}_{\rm rr}(\omega,\bm k=0)={8T\gamma_1 M^2\omega^2\left(\chi_{\rm A}^2+\chi_{\rm A}^4\kappa_{\rm A}^2\omega^2+4\chi_{\rm A}^2\kappa_{\rm A}\gamma_1 M^2\right)\over \chi_{\rm A}^2\omega^2+\left(\chi_{\rm A}^2\kappa_{\rm A}\omega^2+4\gamma_1 M^2\right)^2}\,,
\ee
from which we obtain the Kubo formula for $\gamma_1$,
\be
\gamma_1={1\over 8T}\lim_{\omega\to 0}\lim_{M\to 0}\lim_{\bm k\to 0}{1\over M^2}G^{S_{\rm A}}_{\rm rr}(\omega,\bm k)\,,
\ee
where the order of the two limits, $\omega\to 0$ and $M\to 0$, is important.
The axial current fluctuation correlation functions are
\begin{align}
\langle j_{\rm A}^t(\omega,\bm k) j_{\rm A}^t(-\omega,-\bm k)\rangle&={2T(\lambda_{\rm A}k^2+4\gamma_1 M^2)\left[\chi_{\rm A}^2+\chi_{\rm A}^4\kappa_{\rm A}^2\omega^2+\chi_{\rm A}^2\kappa_{\rm A}(\lambda_{\rm A}k^2+4\gamma_1 M^2)\right]\over \chi_{\rm A}^2\omega^2+\left(\chi_{\rm A}^2\kappa_{\rm A}\omega^2+\lambda_{\rm A}k^2+4\gamma_1 M^2\right)^2}\,, \nonumber\\ \langle j_{\rm A}^i(\omega,\bm k) j_{\rm A}^j(-\omega,-\bm k)\rangle&=
2T\lambda_{\rm A}\left[\delta^{ij}-\hat{k}^i\hat{k}^j{\lambda_{\rm A} k^2(\chi_{\rm A}^2\kappa_{\rm A}\omega^2+\lambda_{\rm A} k^2+4\gamma_1 M^2)\over\chi_{\rm A}^2\omega^2+\left(\chi_{\rm A}^2\kappa_{\rm A}\omega^2+\lambda_{\rm A} k^2+4\gamma_1 M^2\right)^2} \right]\,.
\end{align}
We obtain the Kubo formula for the conductivity as usual,
\be
\lambda_{\rm A}={1\over 2T}\lim_{\omega,\bm k,M\to 0}{\omega^2\over k^2}G^{tt}_{\rm rr}(\omega,\bm k)\,,
\ee
while the Kubo formula for $\kappa_{\rm A}$ in this phase requires a higher order term in momentum expansion, which is understandable since $\kappa_{\rm A}$ in this phase is a higher order effect in gradient expansion.

In the phase where the axial symmetry is spontaneously broken, the pions are propagating modes satisfying the EOM,
\be
(\chi_{\rm A} \partial_t^2-\chi_{\rm A} v_{\rm s}^2{\bm\nabla}^2+\chi_{\rm A} m_\pi^2-\chi_{\rm A}^2\kappa_{\rm A}\partial_t^3-\lambda_{\rm A}{\bm\nabla}^2\partial_t+4\gamma M^2\partial_t )\phi_{\rm r}=\partial_t \xi_{\kappa {\rm A}}+\partial_i \xi_{\rm A}^i-\xi_{\gamma A}\,,
\ee
where $\gamma=\gamma_1+\gamma_2$, $m_\pi^2=4\sigma M/\chi_{\rm A}$ is the pion mass square, and $v_{\rm s}^2$ is the pion velocity square given by (\ref{eq:v-updated}) with $\chi = \chi_{\rm A}$.
The current fluctuation correlation functions are computed to be
\begin{align}
G_{\rm rr}^{tt}(\omega,\bm k)&={2T\chi_{\rm A}^2\left\{(\lambda_{\rm A}k^2+4\gamma M^2)\left[1+\chi_{\rm A}^2\kappa_{\rm A}^2\omega^2+\kappa_{\rm A}(\lambda_{\rm A}k^2+4\gamma M^2)\right]\omega^2+
\chi_{\rm A}^2\kappa_{\rm A}\omega_k^4\right\}\over \chi_{\rm A}^2(\omega^2-\omega_k^2)^2+\left(\chi_{\rm A}^2\kappa_{\rm A}\omega^2+\lambda_{\rm A}k^2+4\gamma M^2\right)^2\omega^2}\,,\\ 
%%%
G_{\rm rr}^{ij}(\omega,\bm k)&=2T\left[\lambda_{\rm A}\delta^{ij}+k^ik^j{(f_\pi^4-\lambda_{\rm A}^2\omega^2)(\chi_{\rm A}^2\kappa_{\rm A}\omega^2+\lambda_{\rm A}k^2+4\gamma M^2)+2\lambda_{\rm A}\chi_{\rm A}f_\pi^2(\omega^2-\omega_k^2)\over \chi_{\rm A}^2(\omega^2-\omega_k^2)^2+\left(\chi_{\rm A}^2\kappa_{\rm A}\omega^2+\lambda_{\rm A}k^2+4\gamma M^2\right)^2\omega^2}\right]\,,\nonumber\\
\end{align}
where $\omega_k^2\equiv v_{\rm s}^2 k^2+m_\pi^2$.
From this, we obtain a simple Kubo formula for $\kappa_{\rm A}$ as
\be
\kappa_{\rm A}={1\over 2T\chi_{\rm A}^2}\lim_{\omega_k \to 0 }\lim_{\omega\to 0} G_{\rm rr}^{tt}(\omega,\bm k)\,.
\ee
Note that $\omega\to 0$ limit must be taken in advance of $k\to 0$ or $M\to 0$ (i.e., $\omega_k\to 0$), where the order of these last two limits does not matter. 
Another Kubo formula can also be obtained using the source fluctuation correlation function.
We have $G_{\rm rr}^{S_{\rm A}}(\omega,\bm k=0)=\omega^2 G_{\rm rr}^{tt}(\omega,\bm k=0)$ by the Ward-Takahashi identity, and we have the Kubo formula for $\kappa_{\rm A}$ in this phase as
\be
\kappa_{\rm A}={1\over 2T\chi_{\rm A}^2}\lim_{\omega\to 0}\lim_{M\to 0}\lim_{\bm k\to 0}{\omega^2\over m_\pi^4}G^{S_{\rm A}}_{\rm rr}(\omega,\bm k)={1\over 32T\sigma^2}\lim_{\omega\to 0}\lim_{M\to 0}\lim_{\bm k\to 0}{\omega^2\over M^2}G^{S_{\rm A}}_{\rm rr}(\omega,\bm k)\,.
\ee
The Kubo formula for $\gamma$ in this phase requires a higher order term in momentum expansion since the effect of $\gamma$ in this phase is sub-leading compared to that of $\kappa_{\rm A}$. 

\section{Conclusion \label{conclude}}

In this paper, we constructed the low-energy effective theory on the SK contour for systems with ${\rm U}(1)$, ${\rm SU}(N)$, and the chiral ${\rm SU}(N)_{\rm L}\times {\rm SU}(N)_{\rm R}$ symmetries, incorporating their small explicit symmetry-breaking parameters in a fully nonlinear and/or non-Abelian manner. 
For each symmetry, we investigated relaxation dynamics both in the symmetry-broken and symmetry-restored phases. 
We carefully analyzed the spurious symmetry transformations associated with explicit symmetry breaking and identified all possible terms that maintain the ``gauge invariance" when the explicit breaking parameters also transform under spurious symmetry transformations.  
We found that the relaxation term in the approximate conservation law should be second-order in the explicit symmetry-breaking parameter, such as the quark mass in QCD for the chiral symmetry.

There are several future directions possibly stemming from our work. 
While our EFT approach to the hydrodynamics of approximate symmetries is a powerful method, there are complemetary theoretical approaches capturing dynamics in the quasi-hydrodynamic regime.
It would be interesting to compare the present results with those from other approaches, e.g., the thermodynamic approach based on the second law of thermodynamics~\cite{Onsager:1931jfa,Landau:Fluid}, the statistical mechanical approaches such as the projection operator method~\cite{Mori:1965oqj,Zwanzig2001,Hayata:2014yga} or the nonequilibrium statistical operator method~\cite{Zubarev:1979afm,Becattini:2014yxa,Hayata:2015lga}, and microscopic approaches based on, e.g., the kinetic theory (see, e.g., Ref.~\cite{Hidaka:2022dmn} for a recent review) or the holography~\cite{Grozdanov:2018fic,Ishigaki:2020vtr,Ammon:2021pyz,Cao:2022csq}.
In these formalisms, the dissipative terms to second order in the quark mass have not been systematically studied.
We hope to address the thermodynamic approach in a forthcoming paper~\cite{future}.
It would also be interesting to take into account the energy-momentum sector that we have neglected in this work.

\vskip 1cm \centerline{\large \bf Acknowledgment} \vskip 0.5cm

This work is supported by the U.S. Department of Energy, Office of Science, Office of Nuclear Physics, grant No. DEFG0201ER41195 (NS, MS, H-UY), and by Japan Society for the Promotion of Science (JSPS) KAKENHI Grant Nos. 22K20369 and 23H01174 (MH).
MH thanks Luca V. Delacr\'{e}taz for useful discussions during the YITP-RIKEN iTHEMS International Molecule-type Workshop (YITP-T-24-04) on “Advances in Fluctuating Hydrodynamics: Bridging the Micro and Macro Scales.”

\bibliographystyle{JHEP}
\bibliography{refs.bib}
\end{document}